\documentclass[useAMS,usenatbib]{mn2e}

\usepackage[active]{srcltx}
\usepackage{graphicx}
\usepackage{txfonts}
\usepackage{natbib}
\usepackage{multirow}
\usepackage{longtable}
\usepackage{subfig}
\usepackage[applemac]{inputenc}
\usepackage{rotating}
\usepackage{tablefootnote}
\usepackage{url}
\usepackage[multiple]{footmisc}

\usepackage{color}
\usepackage{soul}
\definecolor{jaune}{rgb}{1.0, 1.0, 0.0}

\newcommand{\bz}{\ensuremath{\langle B_z\rangle}}

\newcommand{\ddeg}{$^{\circ}$}

\newcommand*{\vcenteredhbox}[1]{\begingroup
\setbox0=\hbox{#1}\parbox{\wd0}{\box0}\endgroup}

\def\gtrsim{\mathrel{\hbox{\rlap{\hbox{\lower4pt\hbox{$\sim$}}}\hbox{$>$}}}}
\def\ltsim{\mathrel{\hbox{\rlap{\hbox{\lower4pt\hbox{$\sim$}}}\hbox{$<$}}}}

\title[The MiMeS Survey: Introduction and Overview]{The MiMeS Survey of Magnetism in Massive Stars: Introduction and overview\thanks{Based on MiMeS Large Program and archival spectropolarimetric observations obtained at the Canada-France-Hawaii Telescope (CFHT) which is operated by the National Research Council of Canada, the Institut National des Sciences de l'Univers (INSU) of the Centre National de la Recherche Scientifique (CNRS) of France, and the University of Hawaii; on MiMeS Large Program and archival observations obtained using the Narval spectropolarimeter at the Observatoire du Pic du Midi (France), which is operated by CNRS/INSU and the University of Toulouse; and on MiMeS Large Program observations acquired using HARPSpol on the ESO 3.6m telescope at La Silla Observatory, Program ID 187.D-0917.}}

\author[MiMeS et al.]{G.A. Wade$^1$, C. Neiner$^2$, E. Alecian$^{3,4,2}$, J.H. Grunhut$^5$, V. Petit$^6$, B. de Batz$^2$,
\newauthor{D.A. Bohlender$^{7}$, D. H. Cohen$^{8}$, H.F. Henrichs$^{9}$, O. Kochukhov$^{10}$, J.D. Landstreet$^{11,12}$,}
\newauthor{N. Manset$^{13}$, F. Martins$^{14}$, S. Mathis$^{15,2}$, M.E. Oksala$^{2}$, S.P. Owocki$^{16}$, }
\newauthor{Th. Rivinius$^{17}$, M.E. Shultz$^{18,17,1}$, J.O. Sundqvist$^{19,16,42,43}$, R.H.D. Townsend$^{20}$,}
\newauthor{A. ud-Doula$^{21}$, J.-C. Bouret$^{22}$, J. Braithwaite$^{23}$, M. Briquet$^{2,24}$\thanks{FRS-FNRS Postdoctoral Researcher, Belgium}, A.C. Carciofi$^{25}$, }
\newauthor{A. David-Uraz$^{18,1}$, C.P. Folsom$^3$, A. W. Fullerton$^{26}$, B. Leroy$^2$,W.L.F. Marcolino$^{27}$, }
\newauthor{A.F.J. Moffat$^{28}$, Y. Naz\'e$^{24}$\thanks{FRS-FNRS Research Associate}, N. St Louis$^{28}$, M. Auri\`ere$^{29,30}$, S. Bagnulo$^{12}$, J.D. Bailey$^{31}$, }
\newauthor{R.H. Barb\'a$^{32}$, A. Blaz\`ere$^2$, T. B\"ohm$^{29,30}$, C. Catala$^{33}$, J.-F. Donati$^{30}$, L. Ferrario$^{34}$,}
\newauthor{D. Harrington$^{35,36,37}$, I.D. Howarth$^{38}$, R. Ignace$^{39}$, L. Kaper$^{9}$, T. L\"uftinger$^{40}$,}
\newauthor{R. Prinja$^{38}$, J.S. Vink$^{12}$, W.W. Weiss$^{40}$, I. Yakunin$^{41}$}\\
\newauthor{(All affiliations are located at the end of the paper.)}}

\begin{document}

\date{Accepted . Received , in original form }

\pagerange{\pageref{firstpage}--\pageref{lastpage}} \pubyear{2002}

\maketitle

\label{firstpage}

\begin{abstract}
The MiMeS project is a large-scale, high resolution, sensitive spectropolarimetric investigation of the magnetic properties of O and early B type stars. Initiated in 2008 and completed in 2013, the project was supported by 3 Large Program allocations, as well as various programs initiated by independent PIs and archival resources. Ultimately, over 4800 circularly polarized spectra of 560 O and B stars were collected with the instruments ESPaDOnS at the Canada-France-Hawaii Telescope, Narval at the T\'elescope Bernard Lyot, and HARPSpol at the European Southern Observatory La Silla 3.6m telescope, making MiMeS by far the largest systematic investigation of massive star magnetism ever undertaken. In this paper, the first in a series reporting the general results of the survey, we introduce the scientific motivation and goals, describe the sample of targets, review the instrumentation and observational techniques used, explain the exposure time calculation designed to provide sensitivity to surface dipole fields above approximately 100~G, discuss the polarimetric performance, stability and uncertainty of the instrumentation, and summarize the previous and forthcoming publications.
 \end{abstract}

\begin{keywords}
Stars : rotation -- Stars: massive -- Instrumentation : spectropolarimetry -- Stars: magnetic fields
\end{keywords}


\section{Introduction}

Magnetic fields are a natural consequence of the dynamic plasmas that constitute a star. Their effects are most dramatically illustrated in the outer layers of the Sun and other cool stars, in which magnetic fields structure and heat the atmosphere, leading to time-variable spots, prominences, flares and winds. This vigorous and ubiquitous magnetic activity results from the conversion of convective and rotational mechanical energy into magnetic energy, generating and sustaining highly structured and variable magnetic fields in their outer envelopes whose properties correlate strongly with stellar mass, age and rotation rate. Although the detailed physics of the complex dynamo mechanism that drives this process is not fully understood, the basic principles are well established \citep[e.g.][]{2009ARA&A..47..333D,2009LRSP....6....4F,2010LRSP....7....3C}. 

Convection is clearly a major contributor to the physics of the dynamo. Classical observational tracers of dynamo activity fade and disappear with increasing effective temperature amongst F-type stars (around $1.5~M_\odot$ on the main sequence), at roughly the conditions predicting the disappearance of energetically-important envelope convection \citep[e.g.][]{2008LRSP....5....2H}. As an expected consequence, the magnetic fields of hotter, higher-mass stars differ significantly from those of cool stars \citep{2009ARA&A..47..333D}: they are detected in only a small fraction of stars \citep[e.g.][]{1968PASP...80..281W, 2007pms..conf...89P}, with strong evidence for the existence of distinct populations of magnetic and non-magnetic stars \citep[e.g.][]{1982ApJ...258..639L,2002A&A...392..637S,2007A&A...475.1053A,2010A&A...523A..40A}.

The known magnetic fields of hot stars are structurally much simpler, and frequently much stronger, than the fields of cool stars \citep[][]{2009ARA&A..47..333D}. The large-scale strength and geometry of the magnetic field are stable, in the rotating stellar reference frame, on timescales of many decades \citep[e.g.][]{2000MNRAS.313..851W,2014MNRAS.440..182S}. Magnetic fields with analogous properties are sometimes observed in evolved intermediate-mass stars \citep[e.g. red giants,][]{2011A&A...534A.139A}, and they are observed in pre-main sequence stars and young main sequence stars of similar masses/temperatures with similar frequencies \citep{2013MNRAS.429.1001A}. Most remarkably, unlike cool stars their characteristics show no clear, systematic correlations with basic stellar properties such as mass \citep{2008A&A...481..465L} or rotation rate \citep[e.g.][]{2000A&A...359..213L,2002A&A...394.1023B}. 


The weight of opinion holds that these puzzling magnetic characteristics reflect a fundamentally different field origin for hot stars than that of cool stars: that the observed fields are not currently generated by dynamos, but rather that they are {\em fossil fields}; i.e. remnants of field accumulated or enhanced during earlier phases of stellar evolution \citep[e.g.][]{1982ARA&A..20..191B,2001ASPC..248..305M,2009ARA&A..47..333D}. In recent years, semi-analytic models and numerical simulations \citep[e.g.][]{2006A&A...450.1077B, 2008MNRAS.386.1947B, 2010A&A...517A..58D, 2010ApJ...724L..34D} have demonstrated the existence of quasi-static large-scale stable equilibrium magnetic field configurations in stellar radiative zones. These solutions bear remarkable qualitative similarities to the observed field characteristics. 


The detailed processes of field accumulation and enhancement needed to explain the characteristics of magnetic fields observed at the surfaces of hot stars are a matter of intense discussion and debate, and range from flux advection during star formation, to protostellar mergers, to pre-main sequence dynamos. Whatever the detailed pathways, due to the supposed relic nature of their magnetic fields, higher-mass stars potentially provide us with a powerful capability: to study how fields evolve throughout the various stages of stellar evolution, and to explore how they influence, and are influenced by, the structural changes that occur during the pre-main sequence, main sequence, and post-main sequence evolutionary phases. 

The first discoveries of magnetic fields in B stars that are sufficiently hot to show evidence of the interaction of the field and the stellar wind occurred in the late 1970s \citep{1978ApJ...224L...5L}. This was followed by the discovery of a small population of magnetic and chemically peculiar mid- to early-B stars \citep{1979ApJ...228..809B,1983ApJS...53..151B}, some of which exhibited similar wind-related phenomena in their optical and/or UV spectra \citep[e.g.][]{1987AJ.....94..737S,1990ApJ...365..665S}. The introduction of new efficient, high resolution spectropolarimeters in the early to mid 2000s led to discoveries of fields in hotter, and frequently chemically normal, B-type stars on the main sequence and pre-main sequence \citep[e.g.][]{2001MNRAS.326.1265D,2006MNRAS.370..629D,2003A&A...406.1019N,2003A&A...411..565N,2008MNRAS.387L..23P,2008MNRAS.385..391A,2008A&A...481L..99A,2013A&A...555A..46H} and in both young and evolved O-type stars \citep{2002MNRAS.333...55D,2006MNRAS.365L...6D}. These discoveries demonstrated that detectable surface magnetism is present in stars as massive as $40-60~M_\odot$.

\begin{figure*}
\begin{centering}
\hspace{-0.9cm}\vcenteredhbox{\includegraphics[width=12.4cm]{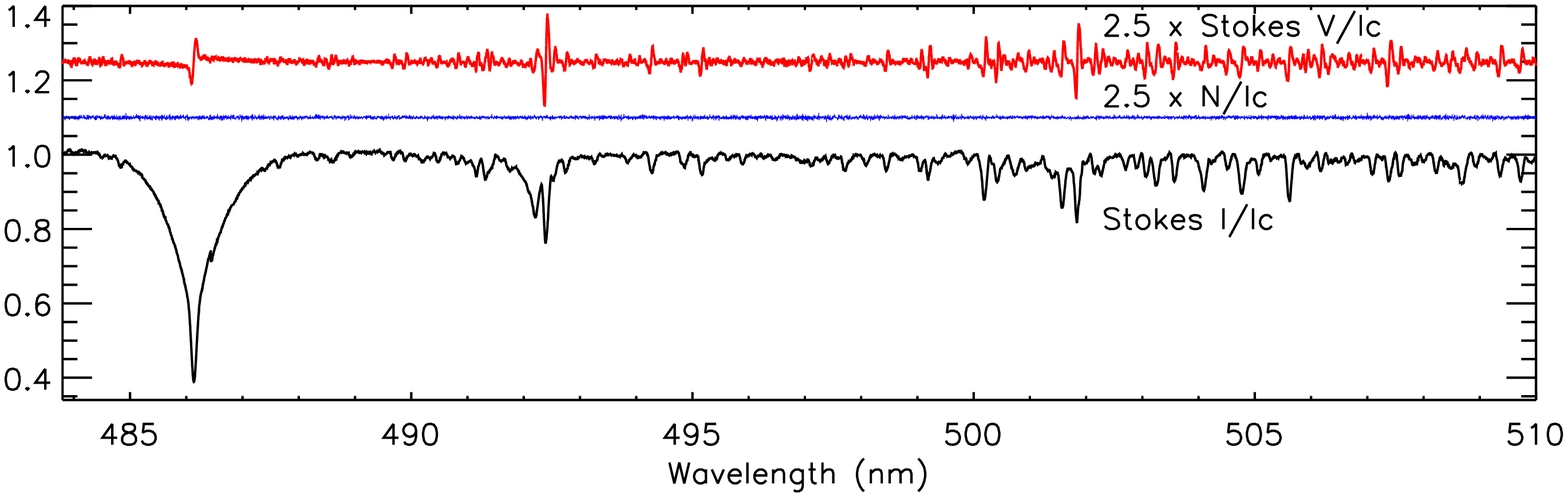}}\hspace{0.25cm}\vcenteredhbox{\includegraphics[width=5.25cm]{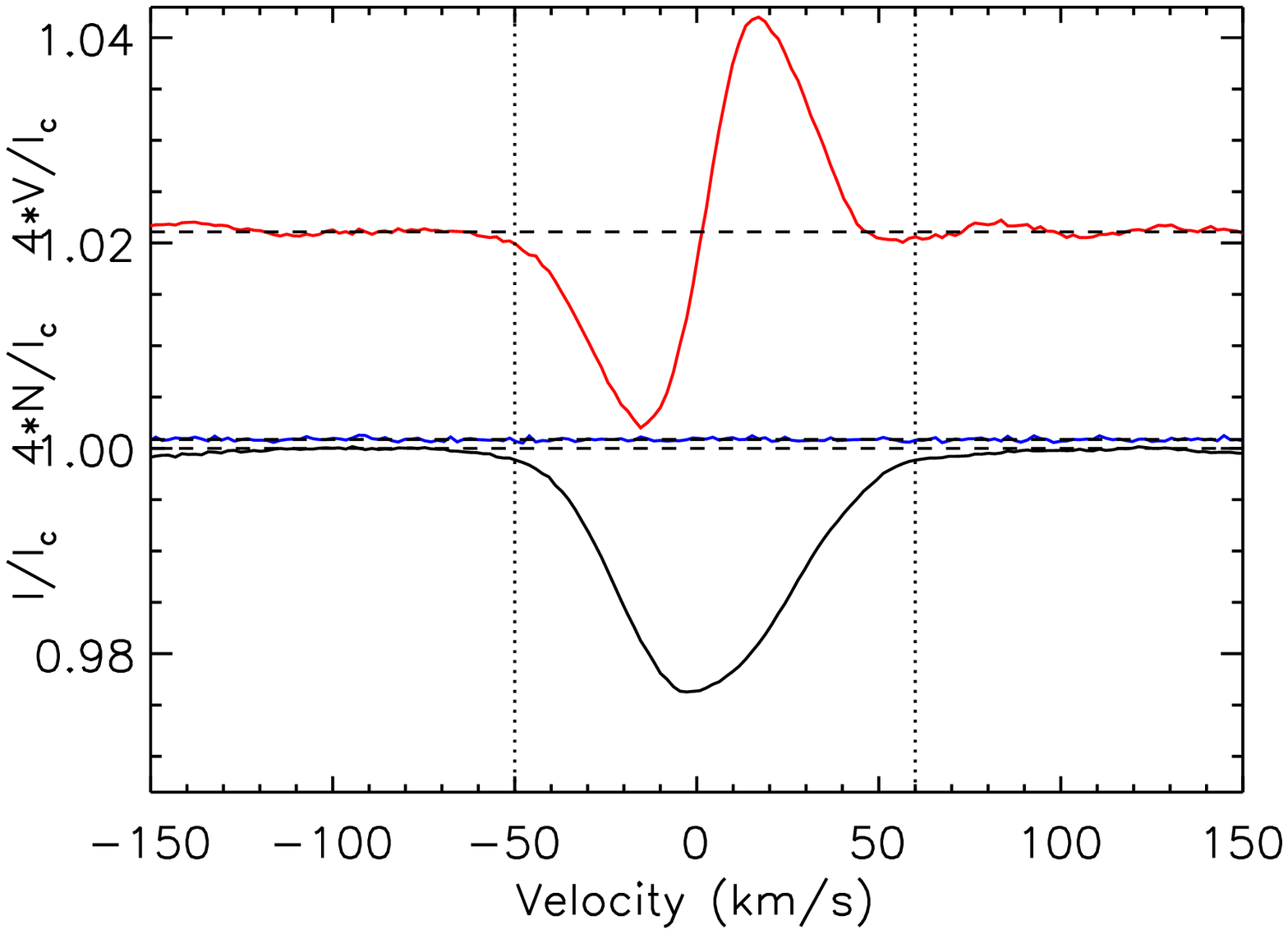}}
\caption{\label{hd175362} {\em Left -}\ A small region of a typical polarized spectrum acquired with the ESPaDOnS instrument during the MiMeS project. This figure illustrates the spectrum of the B5Vp magnetic He weak star HD 175362 (Wolffs' Star), a MiMeS Targeted Component target exhibiting a peak longitudinal magnetic field of over 5~kG. From bottom to top are shown the Stokes $I$ spectrum (in black), the diagnostic null ($N$) spectrum (in blue), and the Stokes $V$ spectrum (in red). Notice the strong polarization variations across spectral lines exhibited in the Stokes $V$ spectrum. Such variations represent signatures of the presence of a strong magnetic field in the line-forming region. The simultaneous absence of any structure in the $N$ spectrum gives confidence that the Stokes $V$ detection is real, and unaffected by significant systematic effects. The $V$ and $N$ spectra have been scaled and shifted vertically for display purposes. The signal-to-noise ratio of this spectrum at 500 nm is 860 per pixel. {\em Right -}\ The Least-Squares Deconvolved (LSD) profiles of the full spectrum corresponding to the left panel. Again, the $V$ and $N$ spectra have been scaled and shifted vertically for display purposes. LSD profiles (described in Sect.~\ref{strategy}) are the principal data product used for diagnosis and measurement of stellar magnetic fields in the MiMeS project. The disc averaged, line of sight (longitudinal) magnetic field measured from the Stokes $V$ profile (between the dashed integration bounds) is $4380\pm 55$~G, while for the $N$ profile it is $-3\pm 16$~G. }
\end{centering}
\end{figure*} 


The Magnetism in Massive Stars (MiMeS) project is aimed at better understanding the magnetic properties of B- and O-type stars through observation, simulation, and theory. The purpose of this paper is to establish the motivation, strategy and goals of the project, to review the instrumentation and observational techniques used (\S\ref{instruments}), to describe the sample of targets that was observed and the exposure time calculations (\S\ref{strategy}), to discuss the polarimetric performance, stability and uncertainty of the instrumentation (\S\ref{performance}), and to summarize the previous and forthcoming publications (\S\ref{summary}). 

\section{Instrumentation and observations}
\label{instruments}

\subsection{Overview}

The central focus of the observational effort of the MiMeS project has been the acquisition of high resolution broadband circular polarization (Stokes $I$ and $V$) spectroscopy. This method relies on the circular polarization induced in magnetically-split spectral line $\sigma$ components due to the longitudinal Zeeman effect \citep[see, e.g.][for details concerning the physical basis of the method]{1989FCPh...13..143M,2004ASSL..307.....L,2009ARA&A..47..333D,2009EAS....39....1L,2009EAS....39...21L}. Although some high resolution linear polarization (Stokes $QU$) and unpolarized spectroscopy has been acquired, the data described in this paper and those that follow in this series, will be primarily Stokes $I+V$ spectra. 

High spectral resolution ($R\gtrsim 65,000$) and demonstrated polarimetric precision and stability were the principal characteristics governing the selection of instrumentation. As a consequence, the project exploited the entire global suite of suitable open-access instruments:  the ESPaDOnS spectropolarimeter at the Canada-France-Hawaii Telescope (CFHT), the Narval instrument at the T\'elescope Bernard Lyot (TBL) at Pic du Midi observatory, and the HARPSpol instrument at ESO's La Silla 3.6m telescope. As demonstrated in previous studies \citep[e.g.][]{2012ApJ...750....2S,2014MNRAS.444..429D,2006MNRAS.370..629D}, these instruments provide the capability to achieve high magnetic precision, to distinguish the detailed contributions to the complex spectra of hot stars, and to construct sophisticated models of the magnetic, chemical and brightness structures of stellar surfaces, as well as their circumstellar environments.

\subsection{Observational strategy}

To initiate the observational component of the MiMeS project, the collaboration was awarded a 640 hour Large Program (LP) with ESPaDOnS. This award was followed by LP allocations with Narval (137 nights, or 1213 hours), and with HARPSpol (30 nights, or 280 hours). 


Some of this observing time was directed to observing known or suspected magnetic hot stars (the MiMeS Targeted Component, `TC'), while the remainder was applied to carrying out a broad and systematic survey of the magnetic properties of bright O and B stars (the Survey Component,`SC'). This allowed us to obtain basic statistical information about the magnetic properties of the overall population of hot, massive stars, while also performing detailed investigations of individual magnetic massive stars. An illustration of a typical MiMeS spectrum of a magnetic TC star is provided in Fig.~\ref{hd175362}.


Most observations were further processed using the Least-Squares Deconvolution (LSD) procedure \citep{1997MNRAS.291..658D,2010A&A...524A...5K}. LSD is a cross-correlation multiline procedure that combines the signal from many spectral lines, increasing the effective signal-to-noise ratio (SNR) of the magnetic field measurement and yielding the highest sensitivity magnetic diagnosis available \citep{2000MNRAS.313..851W,2008A&A...481..465L}. The LSD procedure used in MiMeS data analysis is described in more detailed in Sect.~\ref{LSDsection}.

\begin{figure}
\begin{centering}
\includegraphics[width=8cm]{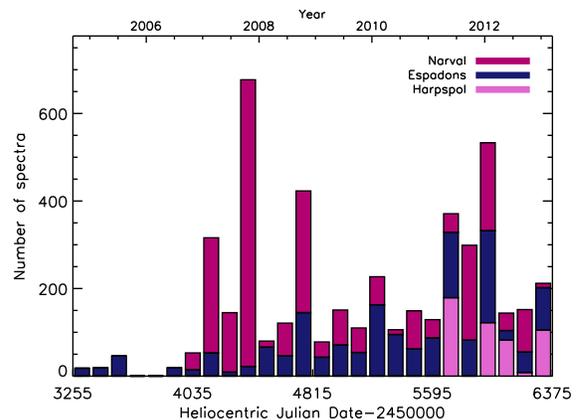}
\caption{\label{datatime}Distribution of data acquisition with time, showing observations acquired with individual instruments (in different colours).}
\end{centering}
\end{figure}

Over 4800 spectropolarimetric observations of 560 stars were collected through LP and archival observations to derive the MiMeS SC and TC. Approximately 50\% of the observations were obtained with Narval, 39\% with ESPaDOnS and the remainder (11\%) with HARPSpol. The distribution of observation acquisition with time is illustrated in Fig.~\ref{datatime}.

\subsection{ESPaDOnS}
\label{SectEspadons}

\subsubsection{Instrument}
\label{espdesc}

ESPaDOnS - the Echelle SpectroPolarimetric Device for the Observation of
Stars \citep[e.g.][]{2003ASPC..307...41D} - is CFHT's optical, high resolution echelle spectrograph and
spectropolarimeter. The instrument consists of a bench-mounted
cross-dispersed echelle spectrograph, fibre-fed from a
Cassegrain-mounted polarimeter unit. The polarimeter unit contains all
instrumentation required for guiding, correction of atmospheric dispersion, calibration
exposures (flat-field, arc and Fabry-Perot frames), and polarimetric analysis. It employs
Fresnel rhombs as fixed quarter-wave and two rotatable half-wave retarders, and
a Wollaston prism as a polarizing beamsplitter. The analyzed starlight
is transported using two long optical fibres to the spectrograph, located in
the CFHT's inner coud\'e room, and housed in a thermal enclosure to minimize
temperature and pressure fluctuations. A fibre agitator is located immediately before the entrance to the spectrograph.
The role of this device is to remove modal noise present in the light transmitted through the optical fibers \citep{2001PASP..113..851B}.

A tunable Bowen-Walraven image slicer slices the twin 1.6" circular images of
the fiber heads at a rate of 3 slices per fiber, producing images of a
pseudo-slit $\sim$12 pixels wide. The slit is tilted, resulting in sampling the pseudo-slit image in such a way 
that resolution is enhanced. As a consequence, the ultimate sampling of one ``spectral bin" is 0.6923 ``CCD bins". 
 
In polarimetry mode, forty spectral orders are captured in each exposure containing both
polarized beams. These curved orders are traced using
flat-field exposures. The images of the pseudo-slit are also tilted with
respect to the detector's rows, and the tilt is measured using 
Fabry-Perot exposures, which produce regularly spaced images of the
pseudo-slit along each order.

\subsubsection{Observations}
\label{Eobs}

A typical spectropolarimetric observation with ESPaDOnS captures a
polarized spectrum in the Stokes $I$ and $V$ parameters with mean
resolving power $R = \lambda/\Delta\lambda\sim65\,000$, spanning the
wavelength range of 369-1048 nm with 3 very small gaps: 922.4-923.4 nm,
960.8-963.6 nm, 1002.6-1007.4 nm. Spectropolarimetric observations of a
target are constructed from a series of 4 individual subexposures,
between which the orientation of the half-wave rhombs is changed, so as
to switch the paths of the orthogonally polarized beams. This allows
the removal of spurious astrophysical, instrumental and
atmospheric artefacts from the polarization spectrum to first
order \citep[e.g.][]{1997MNRAS.291..658D}. For further details on the instrument characteristics and
observing procedure, see \citet{2012MNRAS.426.1003S} and Appendix A of \citet{2013ApJ...764..171D}. 

LP observations of MiMeS targets with ESPaDOnS were initiated in July 2008,
and continued until January 2013. A total of 640 hours were allocated to
the LP (see Table~\ref{allocatedvalidated}). CFHT program identifications associated with MiMeS were P13
(highest priority, about 1/3 of the time awarded) and P14 (lower
priority, about 2/3 of the time awarded) prefixed by the semester ID
(e.g. data acquired during semester 2010B have program ID 10BP13 and
10BP14).  

CFHT observations were conducted under a Queued Service Observing (QSO)
operations scheme. In this scheme, MiMeS observations were scheduled on a
nightly basis,
according to observability of targets, specified time constraints or
monitoring frequencies, weather, and seeing conditions,
in combination with observations requested by other LPs and 
regular observing programs, in order to optimally satisfy the observing requirements
and constraints of the various programs. 

Due to the particular
characteristics of CFHT's 3 primary instruments (ESPaDOnS; a wide-field,
prime focus optical imager MegaCam; and a wide-field, prime focus
infra-red imager WIRCam), only one instrument can be used on the
telescope at a time. As a consequence, ESPaDOnS's fibers are periodically
connected and disconnected from the polarimetric module mounted on the
telescope. The polarimetric module is only removed from the Cassegrain
focus environment when the Adaptive Optics Bonnette is used, on average
once a year, or during engineering shutdowns (e.g. for re-aluminization of the 
primary mirror, which occurred in Aug. 2011). During the 9 semesters of
observation, ESPaDOnS's fibers were typically dis- and re-connected 2-4 times
per semester. The instrument was operational for about 40 total nights per
semester. A total of 726.8 LP hours were observed and 594.2 hours were
{\em validated} on 1519 spectra of 221 targets,
corresponding to a validated-to-allocated ratio of 93\% (see
Table~\ref{allocatedvalidated}). Observations were considered to be validated when
they met the specified technical requirements of the observation, typically minimum SNR and any scheduling requirements
(e.g. for TC targets). However, unless unvalidated observations were patently unusable 
(e.g. incomplete exposure sequences, SNR too low for reduction),
they were usually analysed and included in the analysis. In the case of SC targets,
such observations were often one of several spectra obtained as part of a multi-observation
sequence. In the case of TC targets, such observations (if of lower SNR) might still usefully contribute to sampling
the phase variation of the target, or (if obtained at the wrong phase) might serve to confirm measurements
obtained at other phases.

While the large majority of these 
spectra represent Stokes $I + V$ observations, a small number (about 120)
correspond to Stokes $I + Q$ or Stokes $I + U$ (linear polarization)
observations of about 10 stars. 

Calibration of the instrument (bias, flat-field, wavelength 
calibration and Fabry-Perot exposures) uses a combination of 
thorium/argon and thorium/neon lamps, with all calibrations taken 
at the beginning or end of the night. The arc spectra are used for the primary wavelength 
calibration; telluric lines are then later used to fine-tune the  
wavelength calibration during the reduction process. Filters are used 
to minimise blooming on the chip at the red end of the spectrum. 
Two tungsten lamps are utilised for the flat field frames, with one 
low intensity lamp being used with a red filter and the other lamp 
being higher intensity and used with a blue filter. The 
Fabry-Perot exposure is used to fit the shape and tilt of the pseudo-slit 
created by the image slicer. 

Approximately 85\% of the ESPaDOnS data included in the SC derive from the Large Program. In addition to the LP observations, suitable public data collected
with ESPaDOnS were obtained from the CFHT archive and are included in our analysis. Archival data corresponded to over 350 polarimetric spectra of about 80 additional targets obtained during engineering and Director's time [04BD51, 04BE37, 04BE80, 06BD01] and by PIs (Catala [05AF05, 06AF07, 06BF15, 07BF14], Dougados [07BF16], Landstreet [05AC19, 07BC08], Petit [07AC10, 07BC17], Montmerle [07BF25], Wade [05AC11, 05BC17, 07AC01]). All good-quality Stokes $V$ spectra of O and B stars acquired in archival ESPaDOnS programs up to the
end of semester 2012B were included.



About 50\% of the included ESPaDOnS observations correspond to TC
targets, and 50\% to SC targets.

\subsubsection{Reduction}


ESPaDOnS observations are reduced by CFHT staff using the Upena pipeline
feeding the Libre-ESpRIT reduction package \citep{1997MNRAS.291..658D},
which yields calibrated $I$ and $V$ spectra (or $QU$ linear polarization
spectra) of each star observed. The Libre-ESpRIT package traces the curved
spectral orders and optimally extracts spectra from the tilted slit. Two
diagnostic null spectra called the $N$ spectra, 
computed by combining the four sub-exposures in such a way as to have
real polarization cancel out, are also computed by Libre-ESpRIT (see
\citet{1997MNRAS.291..658D} or \citet{2009PASP..121..993B} for the
definition of the null spectrum). The $N$ spectra represent an
important test of the system for spurious polarization signals that is
applied during every ESPaDOnS spectropolarimetric observation. 
 
The results of the reduction procedure are one-dimensional spectra in
the form of {\sc ascii} tables reporting the wavelength, the Stokes $I$, $V$, ($Q/U$),
and $N$ fluxes, as well as a formal uncertainty, for each spectral
pixel. The standard reduction also subtracts the continuum polarization, as ESPaDOnS only accurately and reliably measures polarization in spectral lines\footnote{Few of the MiMeS targets show significant linear continuum polarization (although WR stars, and to a lesser extent some Be stars, are exceptions). In the standard reduction employed for all non-WR stars, Libre-ESpRIT automatically removes any continuum offset from both $V$ and $N$ using a low-degree order-by-order fit.}. 
CFHT distributes the reduced
polarimetric data in the form of 
{\sc fits} tables containing 4 versions of the reduced data: both
normalized and unnormalized spectra, each with heliocentric radial
velocity (RV) correction applied both using and ignoring the RV content
of telluric lines. In this work we employ only the CFHT unnormalized
spectra. We co-added any successive observations of a target. Then, each reduced SC spectrum was normalized order-by-order
using an interactive {\sc idl} tool specifically optimized to fit the
continuum of these stars. The continuum normalisation is found to be
very reliable in most spectral orders. However, the normalization of those orders containing
Balmer lines is usually not sufficiently accurate for detailed analysis of e.g. Balmer line
wings. While the quality of normalization is sufficient for the magnetic diagnosis, custom normalisation 
is required for more specialized analyses \citep[e.g.][]{2015A&A...575A..34M}.

Archival observations were reduced and normalized in the same manner as SC spectra.

In the case of TC targets, normalization was
often customized to the requirements of the investigation of each star. This 
is also the case for stars or stellar classes with unusual spectra, such as 
WR stars \citep[e.g.][]{2014ApJ...781...73D}.

All CFHT ESPaDOnS data, including MiMeS data and archival data discussed above, can be accessed in raw and reduced form through general queries of the CFHT Science Archive\footnote{www.cadc-ccda.hia-iha.nrc-cnrc.gc.ca/en/cfht} via the Canadian Astronomy Data Centre (CADC)\footnote{www.cadc-ccda.hia-iha.nrc-cnrc.gc.ca}. The PolarBase archive\footnote{polarbase.irap.omp.eu} also hosts an independent archive of most raw and reduced data obtained with ESPaDOnS.

\subsubsection{Issues}

During the 4.5-year term of the LP, two activities have
occurred at the observatory that are important in the context of the
MiMeS observations. 

{\em Identification and elimination of significant ESPaDOnS polarimetric
crosstalk:}\  During the commissioning of ESPaDOnS in 2004 it was found
that the instrument exhibited crosstalk between linear polarization and
circular polarization (and vice versa). Systematic investigation of this
problem resulted in the replacement of the instrument's atmospheric
dispersion corrector (ADC) in the fall of 2009, reducing the crosstalk
below 1\%. Periodic monitoring of the crosstalk confirms that it has
remained stable since 2009. However, higher crosstalk levels were likely
present (with levels as high as 5\%) during the first 3 semesters of
MiMeS LP observations. The absence of any significant impact of
crosstalk on most MiMeS observations is confirmed through long-term
monitoring of TC targets as standards, and is addressed in \S\ref{SectTC}\footnote{In this context, the WR stars 
represent a special case. These stars often have strongly linearly polarized lines and continua, and 
as a consequence crosstalk significantly influenced their Stokes $V$ spectra.
Special analysis procedures were required in order to analyze their magnetic properties \citep{2013ApJ...764..171D, 2014ApJ...781...73D}.}.
The crosstalk evolution and mitigation is described in more
detail by \citet{2010SPIE.7735E.145B} and \citet{2012MNRAS.426.1003S}. 

{\em Change of the ESPaDOnS CCD:}\ Until semester 2011A, ESPaDOnS employed a
grade 1 EEV CCD42-90-1-941 detector with 2K x 4.5K 0.0135 mm square pixels (known as
EEV1 at CFHT). This was replaced in 2011A with a new deep-depletion E2V CCD42-90-1-B32
detector (named Olapa). Olapa has exquisite cosmetics and much less red
fringing than EEV1. Another major difference is that Olapa's quantum
efficiency in the red is about twice as high as with EEV1. Commissioning
experiments by CFHT staff, as well as within the MiMeS project, were
used to confirm that observations acquired before and after the CCD
replacement are in excellent agreement. This is discussed further in \S\ref{SectTC}.

\begin{table}
\caption{ESPaDOnS observations 2008B-2012B. ``Validated" observations are deemed by the observatory
to meet stated SNR, scheduling and other technical requirements. Sometimes, due to observatory QSO requirements,
more hours were observed than were actually allocated. This potentially produced ratios of validated-to-allocated time (Val/Alloc)
greater than 100\%.\label{allocatedvalidated}}
\begin{center}\begin{tabular}{llllllllllllllrrc}
\hline    
ID	&	Allocated	&	Observed	&	Validated	&	Val/Alloc\\
	&	(h)	&	(h)	&	(h)	&	(\%)\\
\hline    
08BP13  &       25.5    &       19.8    &       16.5    &       65\\
08BP14  &       61.6    &       65.5    &       57      &       93\\
09AP13  &       28      &       16.9    &       15.7    &       56\\
09AP14  &       43      &       36.2    &       34.2    &       80\\
09BP13  &       22      &       31.2    &       21.9    &       100\\
09BP14  &       34.1    &       43.2    &       33.9    &       99\\
10AP13  &       28      &       32.7    &       26.4    &       94\\
10AP14  &       43      &       51.6    &       42.8    &       100\\
10BP13  &       24      &       29.7    &       26.6    &       111\\
10BP14  &       36      &       38.8    &       34.6    &       96\\
11AP13  &       28      &       29.7    &       25.5    &       91\\
11AP14  &       43.3    &       48.7    &       38.4    &       89\\
11BP13  &       28      &       23.6    &       22.7    &       81\\
11BP14  &       43.3    &       47.8    &       38.3    &       88\\
12AP13  &       25      &       29.7    &       27.8    &       111\\
12AP14  &       46.3    &       53.7    &       52.7    &       114\\
12BP13  &       24.9    &       54.8    &       31.3    &       126\\
12BP14  &       56      &       73.2    &       47.9    &       86\\
\hline				
Total	&	640	&	726.8	&	594.2	&	93\\
\hline
\end{tabular}
\end{center}
\end{table}


\subsection{Narval}
\label{SectNarval}

\subsubsection{Instrument}

Narval\footnote{spiptbl.bagn.obs-mip.fr/INSTRUMENTATION2} is a near-twin of ESPaDOnS installed at the 2m T\'elescope Bernard Lyot (TBL) in
the French Pyr\'en\'ees. It is composed of a Cassegrain polarimeter unit similar to that
of ESPaDOnS, and a similar spectrograph located in the TBL coud\'e room.

Compared to ESPaDOnS, the instrument was only adapted to the smaller telescope
size. Small differences include the diameter of the entrance pinhole of the Cassegrain
unit (2.8" for Narval, versus 1.6" for ESPaDOnS) and the lack of a fibre agitator. However, the sampling
of the (sliced) pinhole image is identical to that of ESPaDOnS. The CCD used at TBL 
(a back illuminated e2v CCD42-90 with $13.5\mu$m pixels) differs from
that used at CFHT. However, each Narval spectrum also captures 40 spectral orders
covering a similar spectral range (370-1050 nm) with the same resolving power
of about 65\,000.

All other technical characteristics of Narval are effectively identical to those of ESPaDOnS
described in Sect~\ref{espdesc}.


\subsubsection{Observations}
\label{Nobs}

Observations of MiMeS targets with Narval were initiated in March 2009, and
continued until January 2013. A total of 1213 hours were allocated to the MiMeS
program, first in the framework of 3 single-semester programs and then as an LP
for 5 additional semesters (all of these observations are hereinafter considered to be 'LP' observations). 
TBL program identifications associated with MiMeS
were prefixed by the letter "L", followed by the year (e.g. "12" for 2012) and
the semester (1 or 2 for semesters A and B, respectively), then "N" for Narval,
and the ID of the program itself (e.g. "02" in the case of the LP). The MiMeS
Narval runs are thus L091N02, L092N06, L101N11, L102N02, L111N02, L112N02,
L121N02 and L122N02.

Just like at CFHT, TBL observations are conducted under a QSO operations scheme.
The difference, however, is that Narval is the only instrument
available at TBL and thus stays mounted on the telescope all of the time and
Narval observations can occur on any night (except during technical maintenance
periods or closing periods). Calibration spectra (bias, flat-field, wavelength calibration) are obtained at both the beginning 
and end of each observing night.

In total, 1213 hours were allocated and 564.5 hours were validated on approximately 890
polarimetric observations of about 35 targets, corresponding to a
validated-to-allocated ratio of 46.5\% (see Table~\ref{Narvalallocvalid})\footnote{This low validation ratio results principally from very poor weather and an overly optimistic conversion rate of operational hours per night during the first 3 single-semester programs. The mean ratio for the continuing program is much higher, over 90\%.}.
Observations were only validated when they met the requirements of the program: 
observations were generally not validated when taken under very poor sky
conditions or when the requested observing phase was not met. While the large
majority of these spectra represent Stokes $I + V$ observations, a small number
(about 20) correspond to Stokes $I + Q$ or Stokes $I + U$ (linear polarization)
observations of one star (HD\,37776). 

In addition to the MiMeS observations, suitable public data acquired with
Narval were obtained from the TBL archive and are included in our analysis. Archival data corresponded to over 1550 polarimetric spectra of about 60 additional targets (PIs Alecian [L071N03,
L072N07, L081N02, L082N11, L091N01, L092N07], Bouret [L072N05, L081N09, L082N05,
L091N13], Henrichs [L072N02], Neiner [L062N02, L062N05, L062N07, L071N07,
L072N08, L072N09, L081N08]).



About 550 of the included Narval observations correspond to TC
targets (i.e. about 22\%), and the remainder to SC targets.

Approximately 35\% of the Narval data included in the SC derive from the Large Program or the dedicated single-semester programs summarized in Table~\ref{Narvalallocvalid}.

All TBL Narval data, including MiMeS data and archival data discussed above, can be accessed in raw and reduced form through general queries of the TBL Narval Archive\footnote{tblegacy.bagn.obs-mip.fr/narval.html}. Most observations are also available through PolarBase$^{\rm 6}$.

\subsubsection{Reduction}

Similarly to the ESPaDOnS observations, Narval data were reduced at the observatory using the Libre-ESpRIT
reduction package. TBL distributes the reduced polarimetric data in the
form of {\sc ascii} tables containing either normalized or unnormalized
spectra, with heliocentric radial velocity (RV) correction applied both using
and ignoring the RV content of telluric lines. As with the ESPaDOnS data, for SC (and PI) targets we used unnormalized spectra,
co-added any successive observations of a target,
and normalized them order-by-order using an interactive {\sc idl} tool specifically optimized to fit the
continuum of hot stars.  


\subsubsection{Issues}

During the 4-year term of the MiMeS Narval observations, two technical
events occurred at TBL that are important in the context of the project. 


{\em CCD controller issue:}\ From September 23, 2011 to October 4, 2011, abnormally high noise levels were
measured in the data. This was due to an issue with an electronic card in the CCD
controller. The controller was replaced on October 4 and the noise returned to
normal.

{\em Loss of reference of a Fresnel rhomb:}\ In the summers of 2011 and 2012, a loss of positional reference of Fresnel rhomb \#2 of
Narval was diagnosed. This happened randomly but only at high airmass and high
dome temperature. In 2011 the position was only slightly shifted and resulted
in a small decrease in the amplitude of Stokes $V$ signatures. In 2012 however,
the error in position was sometimes larger and resulted in distorted Stokes
$V$ signatures. This technical problem certainly occurred in 2012 on July 12, 15 to 19, 
and September 4, 7, 11, and 14. It probably also occurred in 2011 on August 17, 18, 20 to 22, and 
in 2012 on July 8 to 11, 22 to 24, August 18 to 20, and September 5, 6 and 8. 
The rest of the MiMeS data collected in the summers of 2011 and 2012 appear to be unaffected. Note that this technical problem cannot
create spurious magnetic signatures, but could decrease our ability to detect
weak signatures and does forbid the quantitative interpretation of magnetic
signatures in terms of field strength and configuration.

Since both of these problems were discovered following data acquisition, MiMeS observations obtained during periods affected by these issues were generally validated, and appear as such in Table~\ref{Narvalallocvalid}.

\begin{table}
\caption{Narval observations 2009A-2012B. Observed and validated times include CCD readout times. ``Validated" observations are deemed by the observatory
to meet stated SNR, scheduling and other technical requirements. Sometimes, due to observatory QSO requirements,
more hours were observed than were actually allocated. This sometimes produced ratios of validated-to-allocated time
greater than 100\%.The conversion from nights to hours at TBL has changed with time: for summer nights it was 9h/n in 2009 and 2010 and then $\sim$7h/n in 2011 and 2012; for winter nights it went from 11h/n in 2009, to $\sim$10h/n in 2010 and then 8h/n in 2011 and 2012. \label{Narvalallocvalid}}
\begin{center}\begin{tabular}{llllll}
\hline    
ID	 & \multicolumn{2}{l}{Allocated} & Observed &  Validated & Val/Alloc \\
	 & (n)       & (h)       & (h)	    &  (h)       & (\%)	     \\
\hline    
L091N02  & 24        & 216       &   27.2   &	27.2	 &   12.6    \\
L092N06  & 24        & 264       &   54.0   &	49.9	 &   18.9    \\
L101N11  & 24        & 216       &   27.7   &	27.7	 &   12.8    \\
L102N02  & 13        & 129       &   112.0  &	100.8	 &   78.1    \\
L111N02  & 13        & 90        &   100.1  &	94.5	 &   105.0   \\
L112N02  & 13        & 104       &   101.3  &	87.2	 &   83.9    \\
L121N02  & 13        & 90        &   97.3   &   95.5	 &   106.1   \\
L122N02  & 13        & 104       &   86.7   &   82.6	 &   79.5    \\
\hline	   		   
Total	 & 137       & 1213      &   605.3  &	564.5	 &   46.5     \\
\hline
\end{tabular}
\end{center}
\end{table}

\subsection{HARPSpol}
\label{SectHarps}

\subsubsection{Instrument}

We also used the HARPSpol \citep{2011Msngr.143....7P} polarimetric mode of the HARPS spectrograph \citep{2003Msngr.114...20M} installed on the 3.6m ESO telescope (La Silla Observatory, Chile). The polarimetric module has been integrated into the Cassegrain unit situated below the primary mirror. As with ESPaDOnS and Narval, the Cassegrain unit provides guiding and calibration facilities, and feeds both fibres of HARPS with light of orthogonal polarization states. The polarimeter comprises two sets of polarization optics that can slide on a horizontal rail. Each set of polarimetric optics consists of a polarizing beamsplitter (a modified Glan-Thompson prism) and a rotating super-achromatic half-wave (for linear polarization) or quarter-wave (for circular polarization) plate that converts the polarization of the incoming light into the reference polarization of the beamsplitter. The light beams are injected into fibres of diameter 1" on the sky, which produce images 3.4 pixels in diameter. They feed the spectrograph installed in a high stability vacuum chamber in the telescope's coud\'e room. The spectra are recorded on a mosaic of two 2k$\times$4k EEV CCDs, and are divided into 71 orders (45 on the lower, blue, CCD, and 26 on the upper, red, CCD). 

\subsubsection{Observations}

A typical polarimetric measurement provides simultaneous Stokes $I$ and $V$ echelle spectra with a mean resolving power of $110\,000$, covering a wavelength range from 380 to 690 nm, with a gap between 526 and 534 nm (separating both CCDs). As with ESPaDOnS and Narval, a single polarization measurement is constructed using 4 successive subexposures between which the quarter-wave plate is rotated by 90\ddeg\ starting at 45\ddeg\ (for circular polarization). The calibration spectra (bias, tungsten flat-field, and ThAr wavelength calibration) are systematically obtained at the beginning of each observing night, and in many cases at the end of the night as well.



\begin{table}
\caption{HARPSpol observations during the periods P87-P91. Columns indicate the run ID, run dates, period ID, the number of allocated nights, the estimated number of equivalent operational hours, and the fraction of time useful for observations.\label{harpsrun}}
\begin{center}\begin{tabular}{llllll}
\hline    
run ID		&	Dates				&	P	&	Nights	& Hrs &	Obs 	\\
(187.D-)			&	(local time)			&			&		&	&	time (\%)			\\
\hline    
0917(A)	&	2011 May 21-27	&	87		&	7	& 70	&	67			\\
0917(B)	&	2011 Dec 9-16		&	88		&	8	&64&		100			\\
0917(C)	&	2012 Jul 13- Aug 1	&	89		&	7	& 70&		52			\\
0917(D)	&	2013 Feb 13-20	&	90		&	8		&64&	93			\\
0917(E)	&	2013 Jun 20			&	91		&	1&10&			100			\\
\hline				
Total			&						&			&	31	& 278	&	80			\\
\hline
\end{tabular}
\end{center}
\end{table}

\begin{figure}
\begin{centering}
\includegraphics[width=8cm]{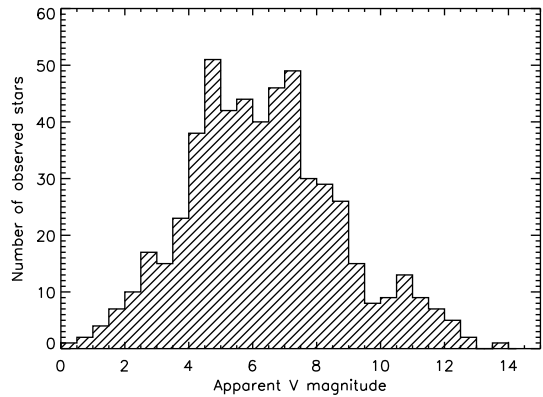}
\caption{\label{mags}Distribution of $V$ band apparent magnitudes of all observed SC targets.}
\end{centering}
\end{figure}

Unlike ESPaDOnS and Narval, HARPSpol is scheduled using a classical scheduling model, in ``visitor" mode. The HARPSpol observations of the MiMeS project were obtained in the framework of Large Program 187.D-0917 over 5 semesters (Periods 87-91, March 2011-September 2013). A total of 30 nights were initially allocated. We obtained one additional night at the end of the project to compensate for bad weather conditions. The observations were obtained during 5 runs, one per Period lasting 1, 7 or 8 nights (Table \ref{harpsrun}). In July 2012 we shared the nights allocated to our run with two other programs, scheduled between July 13th and August 1st. This gave us the possibility to monitor objects with relatively long rotation periods, over more than 7 days, and up to 20 days, allowing us to sample the rotation cycles of TC targets. A total of 532 individual polarized spectra, resulting in 266 coadded observations of 173 stars, were obtained during this LP. All observations were obtained in circular polarization (Stokes $V$) mode. 

All HARPSpol data, including MiMeS data discussed above, can be accessed in raw form through general queries of the ESO Science Archive\footnote{archive.eso.org}.

\subsubsection{Reduction}

The data were reduced using the standard {\sc reduce} package \citep{2002A&A...385.1095P} which performs an optimal extraction of the cross-dispersed echelle spectra after bias subtraction, flat-fielding correction (at which stage the echelle ripple is corrected), and cosmic ray removal. Additionally, we used a set of proprietary {\sc idl} routines developed by O. Kochukhov to perform continuum normalisation, cosmic ray cleaning, and polarimetric demodulation \citep[e.g.][]{2011A&A...536L...6A,2011A&A...525A..97M}. 

The optimally extracted spectra were normalized to the continuum following two successive steps. First, the spectra were corrected for the global response function of the CCD using a heavily smoothed ratio of the solar spectrum measured with HARPSpol, divided by Kurucz's solar flux atlas \citep{1984sfat.book.....K}. The response function corrects the overall wavelength-dependent optical efficiency of the system, the CCD sensitivity (which varies smoothly with wavelength), and also the flux distribution of the flat-field lamp. The latter is not smooth because the HARPS flat-field lamp uses filters to suppress the red part of the spectrum. Then we determined the continuum level by iterative fitting of a smooth, slowly varying function to the envelope of the entire spectrum. Before this final step we carefully inspected each spectrum and removed the strongest and broadest lines (including all Balmer lines, and the strongest He lines), as well as the emission lines, from the fitting procedure.

The polarized spectra and diagnostic null were obtained by combining the four continuum-normalized individual spectra taken at the four different angles of the wave-plate, using the ratio method \citep{1997MNRAS.291..658D}. The spectra of both CCDs, up to this point reduced independently, are then merged to provide a single full spectrum. The heliocentric velocity corrections were computed for the four spectra, and the mean of the four values was applied to Stokes $I$ and $V$, as well as diagnostic null spectra. If successive polarimetric measurements of the same object were obtained, we combined them using a SNR-weighted mean. Each reduced observation was then converted into an {\sc ascii} file in the same format as Narval data. 

All of the HARPSpol data included in the SC derive from the LP, since when the LP was completed no significant archival HARPSpol data existed that were suitable to our purposes.


%
%
%
%
%


\begin{figure}
\begin{centering}
\includegraphics[width=8cm]{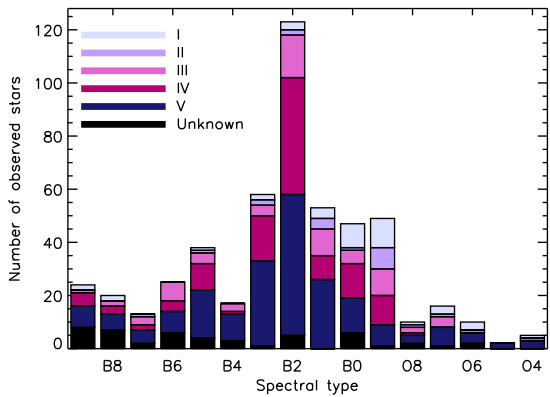}
\caption{\label{STs}Distribution of spectral types and luminosity classes of all observed SC targets, excluding the WR stars, which are discussed in detail by \citet{2014ApJ...781...73D}.}
\end{centering}
\end{figure}

\subsection{Complementary observations}

In addition to the spectropolarimetric data described above, significant complementary data were also acquired, principally in support of the TC. Some of these data were archival in nature (e.g. IUE spectroscopy, Hipparcos photometry, XMM-Newton and Chandra data). Some were acquired through other projects or surveys \citep[e.g. the GOSSS, NoMaDs and OWN surveys,][]{2011ApJS..193...24S, 2012ASPC..465..484M, 2010RMxAC..38...30B} and graciously shared with the MiMeS project through collaborative relationships \citep[e.g.][]{2012MNRAS.425.1278W}. Other data were obtained specifically in support of the MiMeS project: additional Stokes $V$ spectropolarimetry obtained with FORS2, dimaPol and SemPol \citep[e.g.][]{2012MNRAS.419.1610G,2012A&A...545A.119H}, high resolution optical spectroscopy obtained with the FEROS and UVES spectrographs \citep[e.g.][]{2012MNRAS.419.1610G,2012MNRAS.425.1278W,2015MNRAS.449.3945S}, ultraviolet spectroscopy obtained with the Hubble Space Telescope's STIS spectrograph \citep[e.g.][]{2013MNRAS.431.2253M}, X-ray spectroscopy obtained with the XMM-Newton and Chandra X-ray telescopes \citep[e.g.][]{2014ApJS..215...10N,2015arXiv150708621P}, high precision optical photometry obtained with the MOST space telescope \citep{2012MNRAS.419.1610G}, optical phase interferometry obtained with the VLTI \citep{2012AIPC.1429..102R}, optical broadband linear polarization measurements obtained using the IAGPOL polarimeter on the 0.6m telescope of the Pico dos Dias Observatory \citep[e.g.][]{2013ApJ...766L...9C}, and low frequency radio flux measurements \citep{2015MNRAS.452.1245C}.

The details of these observations are described in the respective associated publications.

\section{Targets and exposure times}
\label{strategy}

\begin{table*}
\begin{center}
\caption{MiMeS Targeted Component (TC) sample (32 stars observed, 30 with detected magnetic fields). In addition to HD \# and secondary identifier, we provide the spectral type (typically obtained from the Bright Star Catalogue \citep{1991bsc..book.....H}, the \# of observations acquired and the instruments used (Inst; E=ESPaDOnS, N=Narval, H=HARPSpol), the magnetic field detection status (Det?; T=True, F=False), a reference to completed MiMeS publications, the approximate rotational period, and the peak measured longitudinal field strength. For some TC targets (indicated with a $\dagger$ beside their number of observations), observations include Stokes $Q$/$U$ spectra in addition to Stokes $V$ spectra.\label{TC}}
\begin{tabular}{lllllcllrlllllrrc}
\hline    
HD      &       Other ID        &      Spectral    &       \# obs  & Inst & Det? &  Reference& $P_{\rm rot}$ & $|\bz_{\rm max}|$     \\
    & &   Type    &        &       &        & & (d) & (G)\\
\hline
       &        BD-13 4937     &        B1.5    V     &       31      &EN&T &{\citet{2008A&A...481L..99A}}     & & $1390\pm 395$       \\
  3360 &        $\zeta$~Cas    & B2    IV   &       97      &       N&T & Briquet et al. (to be submitted)  & 5.37 & $30 \pm 5$\\
  34452        &        IQ Aur&  A0p     &40&N&    T & & 2.47 & $1080\pm 310$\\
  35502   && B5      V + A + A& 21 & EN & T &  & 0.85 & $2345 \pm 245$\\
  36485&        $\delta$~Ori C & B2    V      &       10      &N&  T &   & 1.48 & $2460 \pm 85$\\
  36982&        LP Ori&        B0     V      &       15      &      EN&T     &{\citet{2008MNRAS.387L..23P}} & 2.17 & $220 \pm 50$ \\
  37017&        V1046 Ori      & B1.5 V       &       10      &E&     T    & & 0.90 &  $2035 \pm 1075$ \\ 
  37022 &        $\theta^1$~Ori C      &        O7      V      p     &       30      &E&     T & & $15.42$ & $590\pm 115$    \\
  37061&        NU Ori&        B4   &       17      &E&     T        &{\citet{2008MNRAS.387L..23P}} & 0.63 & $310 \pm 50$\\
  37479&        $\sigma$~Ori E & B2    V       p    &       18      & NE&   T &{\citet{2012MNRAS.419..959O}}  & 1.19 & $2345 \pm 55$      \\
  37490&        $\omega$~Ori   & B3    III     e     &       121     &NE&    F & \citet{2012MNRAS.426.2738N} & 1.37 & $\ltsim 90$ \\
  37742 &        $\zeta$~Ori A &        O9.5    Ib    &       495     &N&     T & \citet{2015arXiv150902773B} & 7.0 &$55\pm 15$ \\ 
  37776&        V901 Ori        &      B2      V     &       77$\dagger$      &       EN& T &   & 1.54 & $1310 \pm 65$ \\
  47777&        HD 47777      &        B0.7    IV-V &       10 &E   &  T & {\citet{2014A&A...562A.143F} }&2.64  & $470\pm 85$  \\
  50896       & EZ CMa &        WN4b  &   92$\dagger$&   E&F & \citet{2013ApJ...764..171D} &3.77 & $\ltsim 50$ \\
  64740&HD 64740      & B1.5 V       p        &       17      &HE&    T &  & 1.33 & $660 \pm 60$  \\
  66522&HD 66522      &        B1.5    III     n      &       4&H     &  T & & & $610 \pm 15$  \\
  79158&        36 Lyn& B8    III     pMn  &       29      &N&     T &   & 3.84 & $875\pm 70$    \\
  96446&        V430 Car       &        B2    IV/V          &       10      &H&     T & {\citet{2012A&A...546A..44N}} & 0.85 & $2140\pm 270$ \\
  101412        &        V1052 Cen      &        B0.5    V     &       7       &H&     T  & & 42.08& $785\pm 55$            \\
  124224     & CU Vir & Bp Si & 24 & EN &T&{\citet{2014A&A...565A..83K}} & 0.52 & $940\pm 90$\\
  125823        &        a Cen & B7    III     pv    &       26      &EH&    T     & & 8.82 & $470 \pm 15$  \\
   133880	& HR Lup & 	B8IVp Si 4200	&	2	&	E & T & \citet{2012MNRAS.423..328B} & 0.88 & $4440\pm 160$	\\						
  149438        &        $\tau$~Sco   & B0    V     &       12      &       E & T & & 41.03 & $90\pm 5$ \\
  163472        &        V2052 Oph      & B2   IV-V  &       44      &N&     T &{\citet{2012A&A...537A.148N}}  & 3.64 &  $125 \pm 20$     \\
  175362        &        V686 CrA     & B5    V      p    &       64$\dagger$      &E&     T  & & 3.67 &  $5230 \pm 380$   \\
  184927        &        V1671 Cyg      &        B0.5     IV     nn     &       35$\dagger$      &E&     T &\citet{2015MNRAS.447.1418Y} & 9.53 & $1215 \pm 20$ \\
  191612        &     &        O8    f?p var        &       21      & EN &  T &\citet{2011MNRAS.416.3160W}  & 537 & $ 585 \pm 80$\\
  200775        &        V380 Cep       &        B9   &       63      &NE&    T &{\citet{2008MNRAS.385..391A}}   & 4.33 & $405\pm 80$     \\
  205021        &        $\beta$~Cep    & B1    IV     &       60      &NE&    T &   & 12.00 & $110 \pm 5$  \\
  208057        &        16 Peg & B3   V       e     &       60      &NE&    T &  & 1.37 & $210 \pm 50$ \\
  259135        &        BD+04 1299     &        HBe  &       8 &EN   &T &{\citet{2008A&A...481L..99A}} & & $550\pm 70$ \\
  \hline
\end{tabular}
\end{center}
\end{table*}

\subsection{Targeted Component}

The Targeted Component (TC) was developed to provide high-quality spectropolarimetric data to map the magnetic fields and investigate related phenomena and physical characteristics of a sample of magnetic stars of great interest, at the highest level of sophistication possible for each star. Thirty-two TC targets were identified to allow the investigation of a variety of physical phenomena. The TC sample (summarized in Table~\ref{TC}) consists of stars that were established or suspected to be magnetic at the beginning of the project. The majority of these stars are confirmed periodic variables with periods ranging from approximately 1~d to 1.5 years, with the majority having a period of less than 10 days so that they are suitable candidates for observational monitoring and mapping. They are established to have, or show evidence for, organized surface magnetic fields with measured longitudinal field strengths of tens to thousands of gauss. These targets were typically observed over many semesters, gradually building up phase coverage according to their periods and the operation schedule of the respective instruments. The strict periodicity required for such an observing strategy represents an assumption capable of being tested by the data; this is described in more detail in \S\ref{performance}. Depending on the level of sophistication of the planned analysis (which itself is a function of the limiting quality of the data and individual stellar and magnetic field properties) typically 10-25 observations were acquired for individual TC targets. For a small number of targets, linear polarization Stokes $Q$ and $U$ spectra were also acquired. While the monitoring of the majority of individual TC targets was carried out with a single instrument, for some targets a significant number of observations was acquired with multiple instruments. For example, HD 37940 \citep[$\omega$~Ori;][]{2012MNRAS.426.2738N} and HD 37776 (Landstreet's star; Shultz et al., in prep., Kochukhov et al., in prep.) were extensively observed using both ESPaDOnS and Narval. For some TC targets, only a small number of observations was acquired, either because the target was found to be non-magnetic, non-variable or poorly suited to detailed modelling.


Overall, somewhat less than half of the LP observing time was devoted to observations of the TC. It should be noted that TC targets are the focus of dedicated papers, and are discussed here only for completeness and as a comparison sample.

\begin{figure}
\begin{centering}
\includegraphics[width=8cm]{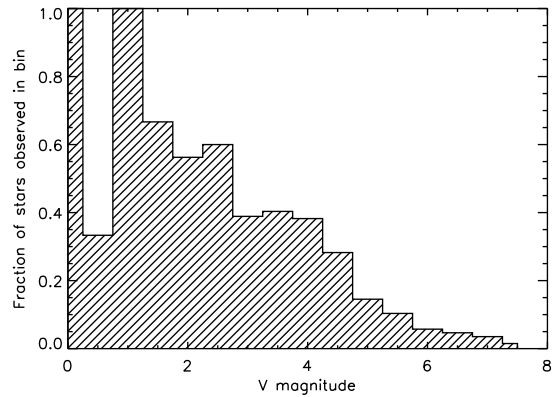}
\caption{\label{completeness1}Completeness of the SC sample as a function of apparent magnitude, to $V=8.0$. This figure illustrates the number of stars observed in the MiMeS survey versus the total number of OB stars of apparent magnitude $V<8.0$, as catalogued by {\sc simbad} \citep{2000A&AS..143....9W}.}
\end{centering}
\end{figure} 

\begin{figure}
\begin{centering}
\includegraphics[width=8cm]{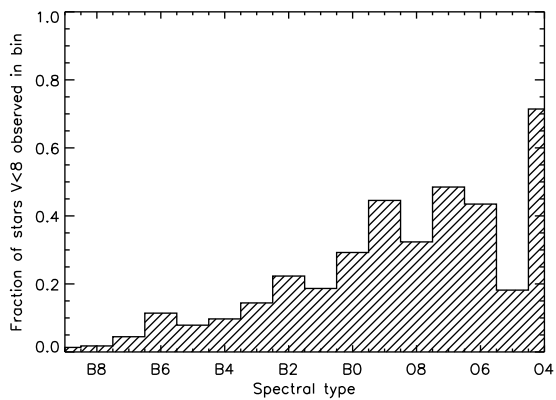}
\caption{\label{completeness2}Completeness of the sample as a function of spectral type for stars with $V<8.0$. This figure \citep[which excludes the WR stars, which are discussed in detail by][]{2014ApJ...781...73D} illustrates the number of stars observed in the MiMeS survey versus the total number of stars of a given spectral type with apparent magnitudes brighter than $V=8.0$.}
\end{centering}
\end{figure} 

\subsection{Survey Component}

The Survey Component (SC) was developed to provide critical missing information about field incidence and statistical field properties for a much larger sample of massive stars, and to provide a broader physical context for interpretation of the results of the TC.  Principal aims of the SC investigation are to measure the bulk incidence of magnetic massive stars, estimate the variation of field incidence with quantities such as spectral type and mass, estimate the dependence of incidence on age, environment and binarity, sample the distribution of field strengths and geometries, and derive the general statistical relationships between magnetic field properties and spectral characteristics, X-ray emission, wind properties, rotation, variability and surface chemistry diagnostics.

The SC sample is best described as an incomplete, principally magnitude-limited stellar sample. The sample is comprised of two groups of stars, selected in different ways. About 80\% of the sample correspond to stars that were observed in the context of the LPs. These stars were broadly selected for sensitivity to surface magnetic fields, hence brighter stars with lower projected rotational velocities were prioritized. To identify this sample, we started from the list of all stars with spectral types earlier than B4 in the {\sc simbad} database. Each target was assigned a priority score according to their apparent magnitude (higher score for brighter stars), $v\sin i$ (higher score for stars with $v\sin i$ below 150 km/s), special observational or physical characteristics (e.g. Be stars, pulsating variables, stars in open clusters) and the existence of UV spectral data, e.g. from the International Ultraviolet Explorer (IUE) archive. These targets were the subject of specific exposure time calculations according to the exposure time model described below. Spectra of the remaining 20\% of the sample were retrieved from the ESPaDOnS and Narval archives. These spectra, corresponding to all stars of spectral types O and B present in the archives at the end of the LPs, were acquired within the context of various programs (generally) unrelated to the MiMeS project (see Sects. \ref{Eobs} and \ref{Nobs}).

\subsection{Properties of observed sample}

\begin{figure*}
\begin{centering}
\includegraphics[width=8cm]{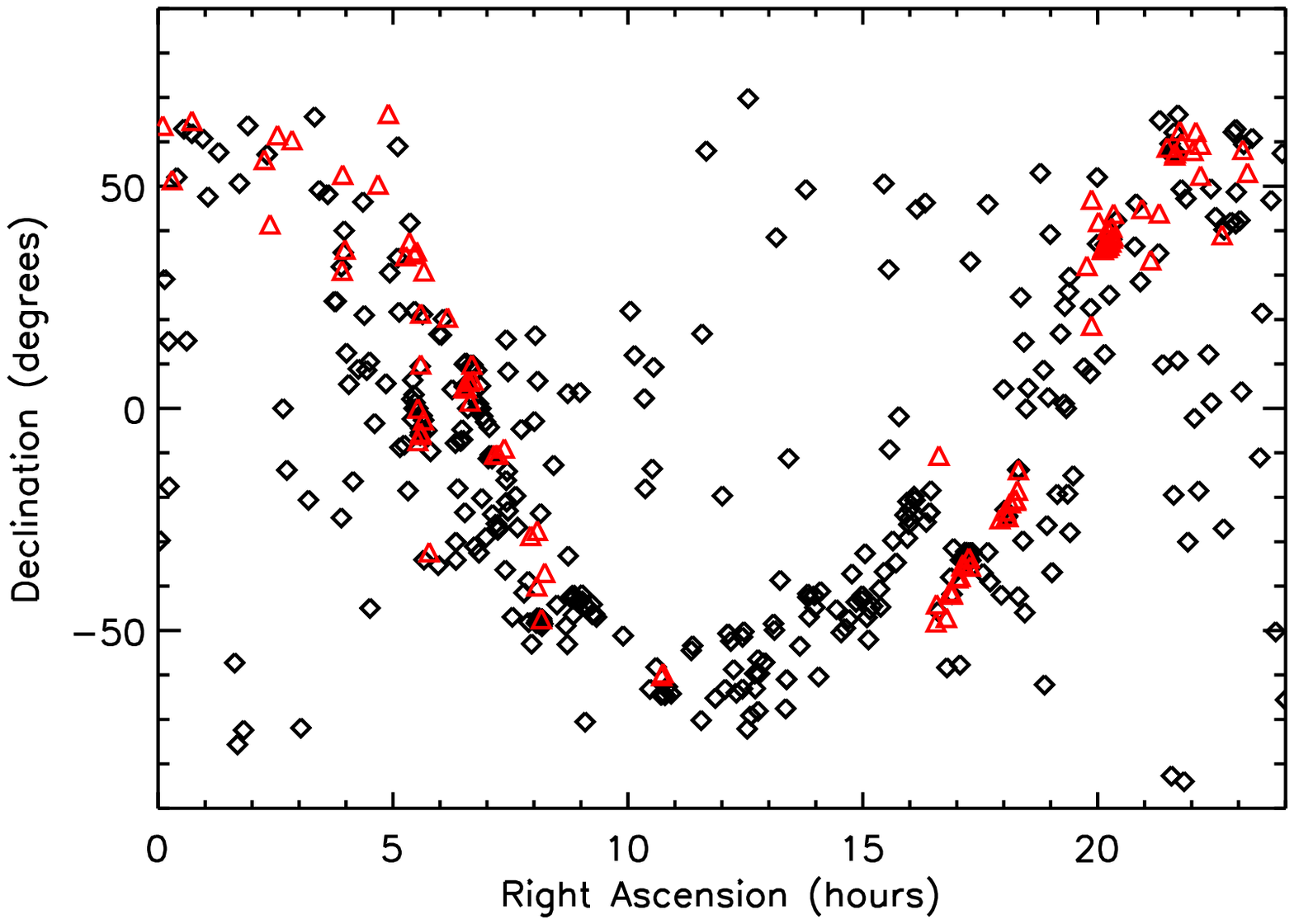}\hspace{1cm}\includegraphics[width=8cm]{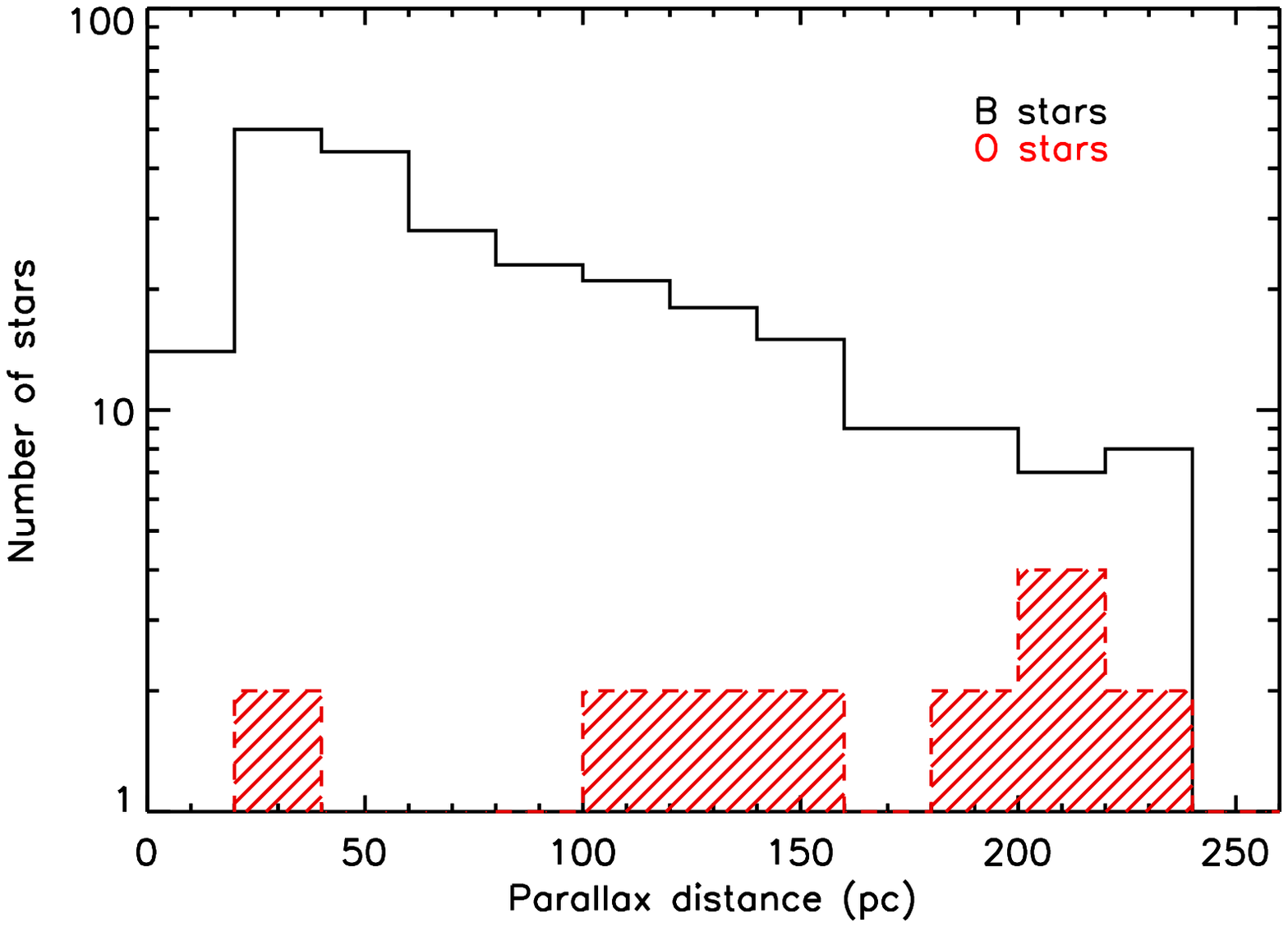}
\caption{\label{distances}{\em Left -}\ Location of all O-type (red triangles) and B-type (black diamonds) SC targets as a function of right ascension and declination. {\em Right -}\ Histogram illustrating distances to the 267 SC stars having high quality ($\pi/\sigma_\pi>4$) measured Hipparcos parallaxes. }
\end{centering}
\end{figure*}

A total of 560 distinct stars or stellar systems were observed (some with more than one instrument), of which 32 were TC targets. Of the 528 SC targets, 106 were O stars or WR stars, and 422 were B stars. Roughly 50\% of the targets were observed with ESPaDOnS, 17\% with Narval and the remaining 33\% with HARPSpol.

We emphasize that our selection process was based on spectral types and motivated by our principal aim: to build a suitable sample to study both the statistics of fossil fields in high-mass stars, and the various impacts of magnetic fields on stellar structure, environment and evolution.


Within this sample, significant subsamples of O and B supergiants, Oe/Be stars and pulsating B stars exist. A systematic survey of the O and B star members of 7 open clusters and OB associations of various ages was also conducted, in order to investigate the temporal evolution of magnetic fields. The cluster sample was selected to include clusters containing very young ($\sim 5$~Myr) to relatively evolved ($\sim 100$~Myr) O and B type stars. These subsamples will be the subjects of dedicated analyses (see \S\ref{summary}). 



%
%
%
%
%



Fig.~\ref{mags} shows that the distribution of apparent magnitude of the sample peaks between 4.5 and 7.5 (the median $V$ magnitude of the sample is 6.2), with an extended tail to stars as faint as 13.6. 

The distributions of $V$ magnitude, spectral type and luminosity class of the SC sample are illustrated in Figs.~\ref{mags} and \ref{STs}, respectively. Approximately 6180 stars of spectral types O and B with apparent magnitudes $V$ brighter than $8.0$ are included in the {\sc simbad} database \citep{2000A&AS..143....9W}; the MiMeS project observed or collected observations of 410 stars brighter than this threshold. Thus overall, we observed about 7\% of the brightest O and B stars. Within this magnitude range, the brightest stars were observed preferentially; for example, 50\% of O and B stars brighter than $V=4$ are included in the sample, and a little more than 20\% of O and B stars brighter than $V=6$ are included. The completeness of the sample as a function of apparent magnitude is illustrated in Fig.~\ref{completeness1}.

Although the initial survey excluded stars with spectral types later than B3, Fig.~\ref{STs} shows that with the inclusion of archival data and due to reclassification of some of our original targets, a significant number of later B-type stars (about 140) form part of the analyzed sample. Including these cooler stars is valuable, since it helps to bridge the gap with the statistics known at later spectral types \citep[F5-B8; e.g.][]{1968PASP...80..281W,2008CoSka..38..443P} and to understand the uncertainties on the spectral types, especially for chemically peculiar stars, for which chemical peculiarities could lead to inaccurate inference of the effective temperature. The strong peak at spectral types B2-B3 reflects the natural frequency of this classification \citep[see, e.g.][]{1991bsc..book.....H}. Spectral types up to O4, as well as a dozen WR stars, are included in the sample. The large majority of SC targets (about 70\%) are main sequence (i.e. luminosity class V and IV) stars. Of the evolved targets, 15\% are giants (class III), 5\% are bright giants (class II) and 10\% are supergiants (class I).

The MiMeS sample preferentially included stars with earlier spectral types. Numerically, the most prominent spectral type observed in the survey was B2 (see Fig.~\ref{STs}). However, as a fraction of all stars brighter than $V=8$, the most complete spectral type was O4 (at 5/7 stars, for 71\%) followed by O7 (at 16/33, for 49\%). Between spectral types of B3 and O4, we observed just under one-quarter (23\%) of all stars in the sky with $V<8$. The completeness as a function of spectral type is illustrated in Fig.~\ref{completeness2}.

To characterize the spatial distribution of the SC targets, we use their equatorial coordinates along with distances for those stars with good-quality parallax measurements. As could be expected, the SC sample is confined primarily to the Galactic plane, as illustrated in Fig.~\ref{distances} (left panel). All of the O-type stars are located in, or close to, the Galactic plane. A majority ($\sim85$\%) of the B-type stars are located close to the Galactic plane. However, a fraction (about 15\%) of the B-type targets are located well away from the plane of the Galaxy. 

We have computed distances to all stars with measured Hipparcos parallaxes significant to $4\sigma$. Amongst the 106 O-type SC stars, only 19 stars have parallaxes measured to this precision. The B-type SC sample of 422 stars, on the other hand, contains 248 stars with precise parallaxes. About 140 of the B stars (more than 1/2 of those with precisely-known parallaxes) are located within 80 pc of the Sun. Approximately one-half of the SC targets have precisely determined parallax distances, and are located within about 250 pc of the Sun. The other half of the sample have poorly determined parallax distances. Overall, it can be concluded that the B-type sample is largely local, whereas the O-type sample is distributed over a larger (but poorly characterized) volume. The distributions of SC target distances are illustrated in Fig.~\ref{distances} (right panel).

As a consequence of the various origins, complicated selection process and diverse properties of the stars included in the SC, the MiMeS sample is statistically complex. An understanding of the ability of the SC to allow broader conclusions to be drawn about the component subsamples will require a careful examination of the statistical properties. This will be the subject of forthcoming papers.

The details of individual stars included in the MiMeS SC sample are reported in Tables~\ref{SCO} and \ref{SCB}. Johnson $V$ magnitudes are from the {\sc simbad} database \citep{2000A&AS..143....9W}. Spectral types for all stars in Tables~\ref{SCO} and \ref{SCB} were obtained from classifications published in the literature or from secondary sources (e.g. estimated from effective temperatures) when unavailable. All sources are cited in the respective tables. 

Targets of the SC sample detected as magnetic were normally scheduled for systematic monitoring, in the same manner as performed for the TC targets. Many such stars have been the subjects of dedicated analyses published in the refereed literature (see \S\ref{summary}).

\subsection{Least-Squares Deconvolution}
\label{LSDsection}

The basic data product employed to evaluate the presence or absence of a magnetic field, and to characterize the field strength or its upper limit, were Stokes $I$, $V$ and diagnostic null $N$ LSD profiles. LSD was applied to all LP and archival spectra, except those of the WR stars \citep[see][for more information concerning analysis of WR stars]{2013ApJ...764..171D,2014ApJ...781...73D}. LSD  \citep[see][]{1997MNRAS.291..658D} is a multiline deconvolution method that models the stellar Stokes $I$ and $V$ spectra as the convolution of a {\em mean profile} (often called the ``LSD profile") with a {\em line mask} describing the wavelengths, unbroadened depths and Land\'e factors of lines occurring in the star's spectrum. The MiMeS LSD procedure involved development of custom line masks optimized for each star, using spectral line data acquired using {\sc extract stellar} requests to the Vienna Atomic Line Database \citep[VALD; ][]{1995A&AS..112..525P}. The LSD codes of both \citet{1997MNRAS.291..658D} and \citet{2010A&A...524A...5K} were normally employed to extract mean profiles. The principal advantage of LSD is that it provides a single set of pseudo line profiles characterizing each spectrum, coherently combining the signal contained in many spectral lines. This yields an easily interpreted, high precision diagnosis of the stellar magnetic field.

The details of the LSD analysis as applied to particular subsamples of the SC and TC are described in published and forthcoming papers.

\begin{figure*}
\begin{centering}
\includegraphics[width=8cm]{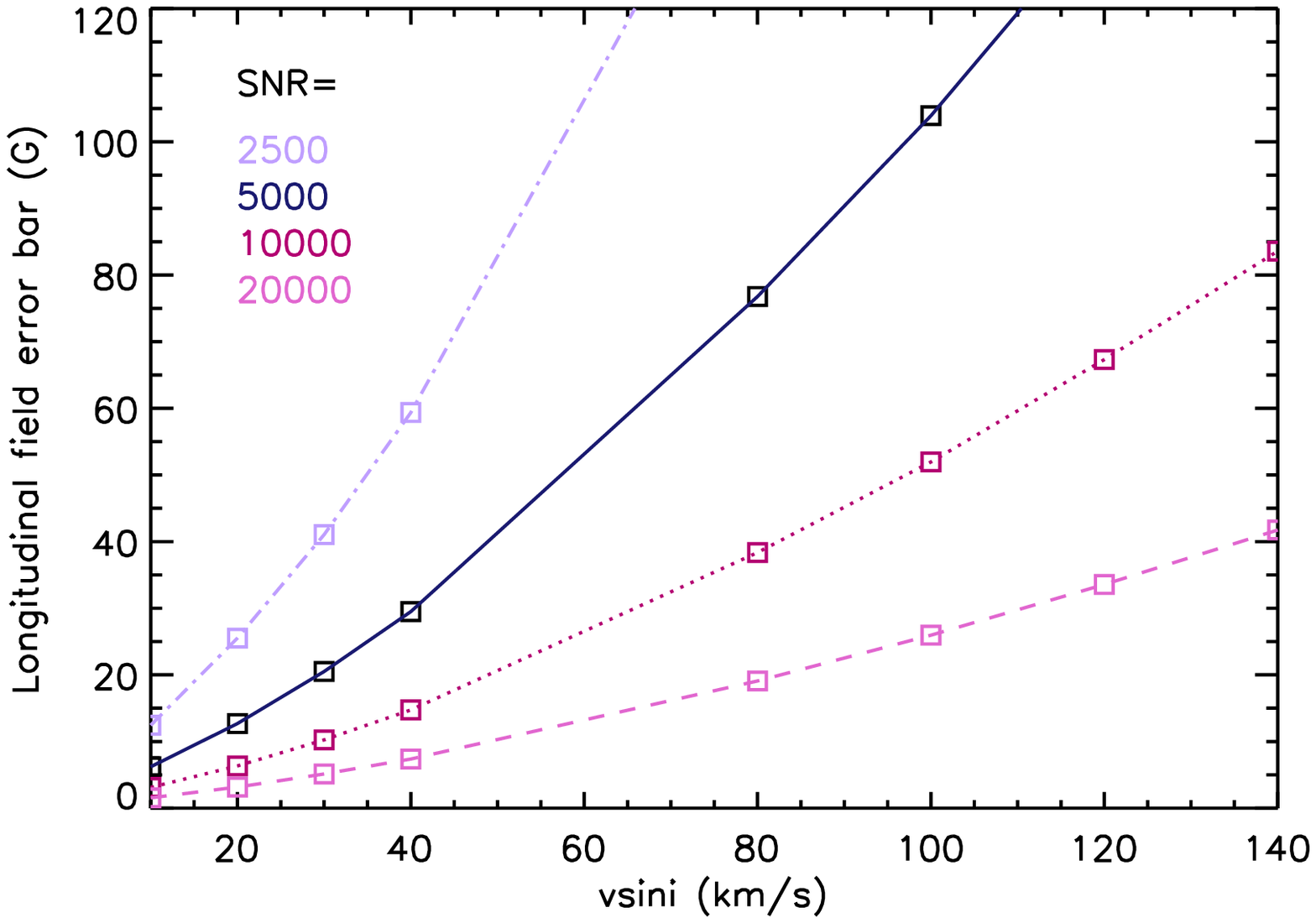}\hspace{1cm}\includegraphics[width=8cm]{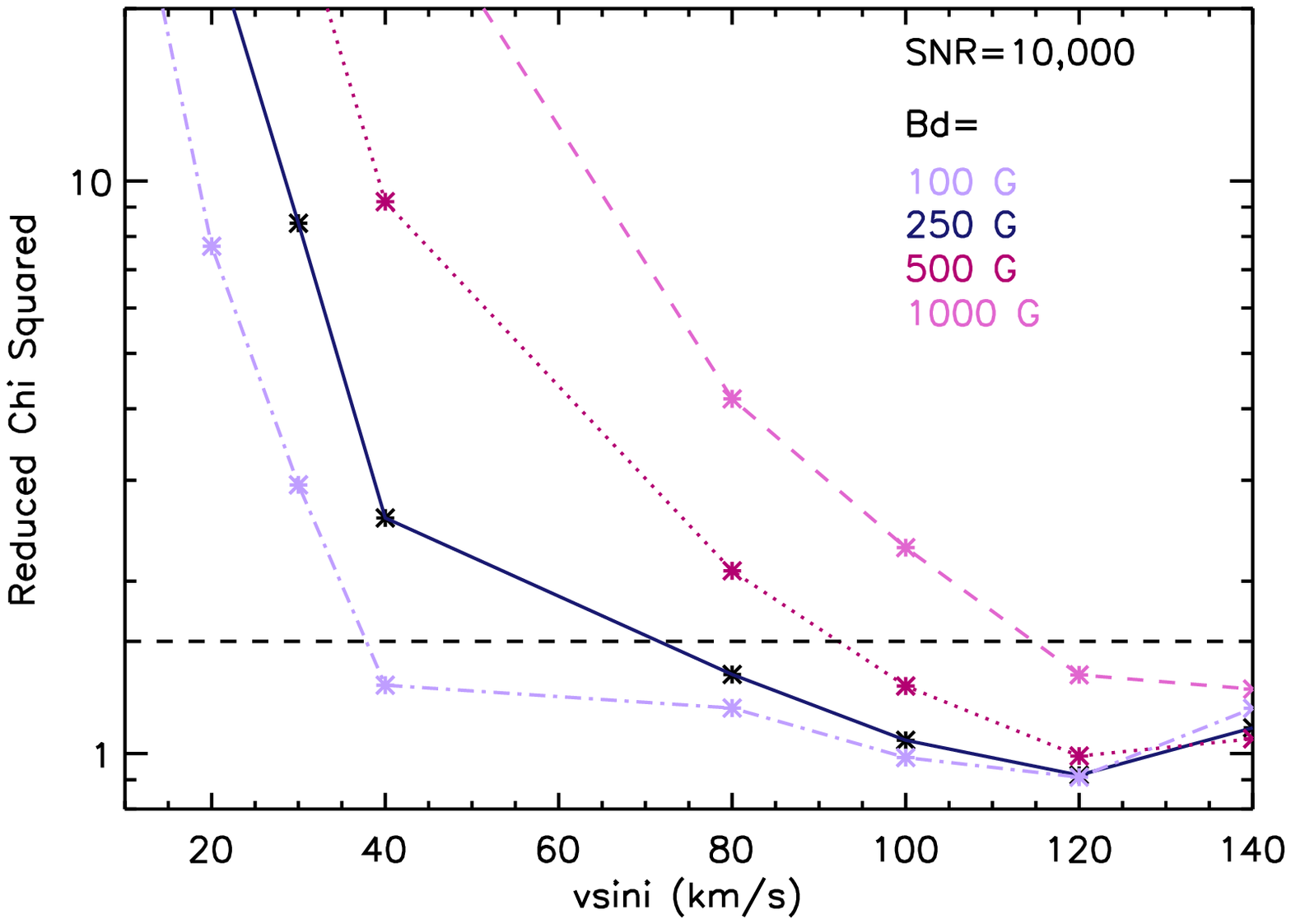}
\caption{\label{exptimemodel}Illustration of exposure model predictions for dipolar magnetic fields. {\em Left -}\ Predicted {\em longitudinal magnetic field} formal uncertainty versus projected rotational velocity, for 4 different SNRs of the LSD profile (2500, 5000, 10000 and 20000). The model predicts a $\sim 50$~G error bar at 100~km/s for an LSD SNR of 10000. {\em Right -}\ Reduced $\chi^2$ of Stokes $V$ within the bounds of the line profile versus $v\sin i$, as a function of {\em surface dipole polar field strength} for an LSD profile SNR of 10000. The dashed line indicates the reduced $\chi^2$ corresponding to a detection at 99.999\% confidence \citep[i.e. a definite detection according to the criteria of][]{1997MNRAS.291..658D}. The weakest fields are detectable only in those stars with relatively sharp lines (e.g. $v\sin i\leq 40$~km/s for 100~G, at this LSD SNR), whereas only stronger fields are detectable in rapidly rotating stars (e.g. 1~kG fields are detectable in stars with $v\sin i\leq 120$~km/s, at this LSD SNR). Different colours and linestyles are used to distinguish the various models.}
\end{centering}
\end{figure*}

\begin{figure}
\begin{centering}
\includegraphics[width=8cm]{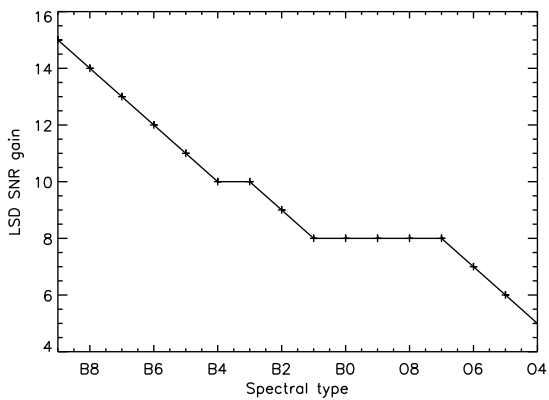}
\caption{\label{gain}Multiplicative gain in SNR versus spectral type assumed in the MiMeS exposure time model.}
\end{centering}
\end{figure}

\subsection{Exposure durations and time budget}
\label{magsens}

LP exposure times were estimated in several ways, as follows.

For TC stars, exposure times typically were based on known amplitudes of Stokes $V$ (or $Q$/$U$) signatures, or estimated based on published field strengths and spectral characteristics. For those TC targets identified as potentially suitable for modelling using individual spectral line Stokes $IV$/$IVQU$ profiles, SNRs per spectral pixel in the reduced spectrum greater than 500 were normally desired. 

The exposure times for SC targets observed within the context of the LPs were computed so as to achieve SNRs corresponding to particular levels of magnetic sensitivity. For the purposes of the survey, ``magnetic sensitivity" was defined in terms of the weakest surface dipole field strength likely to be detected in a particular observation. Such an estimate is rather challenging to make, since it is a function not only of the observational parameters of a star (apparent magnitude, spectral type, line width), but also of the geometry of the surface magnetic field, as well as the assumed rotational phase at the time of observation \citep[e.g.][]{2012MNRAS.420..773P}. Our approach was based on the results of simulations in which Stokes $V$ LSD profiles of a single representative spectral line (selected to be representative of an LSD profile) were synthesized \citep[using the Zeeman code;][]{1988ApJ...326..967L,2001A&A...374..265W} for a large grid of line parameters (depth, $v\sin i$), field geometries and noise levels, with the ultimate aim of deriving an estimate of sensitivity as a function of SNR and $v\sin i$. For some targets no $v\sin i$ was available, and in these cases we assumed a nominal $v\sin i$ of 150~km/s. Illustrative results of these calculations are shown in Fig.~\ref{exptimemodel}.

Because the number of spectral lines present in the stellar spectrum varies significantly with spectral type, the multiplex advantage offered by LSD is also a strong function of this quantity. To quantify the improvement in magnetic precision resulting from LSD, we employed existing spectra of magnetic and non-magnetic stars to estimate the multiplicative gain in SNR $G({\rm ST})$ achieved by application of LSD as a function of spectral type (ST). The gain factor is approximate, with significant variation at each spectral type depending on individual stellar spectral properties. Typically, gain factors exhibit greatest uncertainty at earlier spectral types. A quantitative evaluation of the estimated gain factors, and the overall accuracy of the exposure time model, will be presented in future papers. The gain factors employed in the exposure time calculations are illustrated in Fig.~\ref{gain}.

In order to detect the field strengths of interest ($\sim100-1000$~G), very high SNRs, of order 10000 per spectral pixel in the Stokes $V$ spectrum, were required. Such high SNRs are achievable in two ways: either by co-addition of a series of deep exposures, or by line co-addition using LSD. Often, both of these approaches were combined in order to reach the desired sensitivity. 

Ultimately, surface dipole sensitivity bins of $B_{\rm d}=100,$ 250, 500~G and 1~kG were adopted for the LP survey targets, based principally on published reports of the magnetic strengths of known B- and O-type stars. We implicitly assumed that very strong magnetic fields (with $B_{\rm d}\gg 1$~kG) would be quite rare, whereas weaker fields could be more numerous. 

\begin{figure*}
\begin{centering}
\includegraphics[width=8cm]{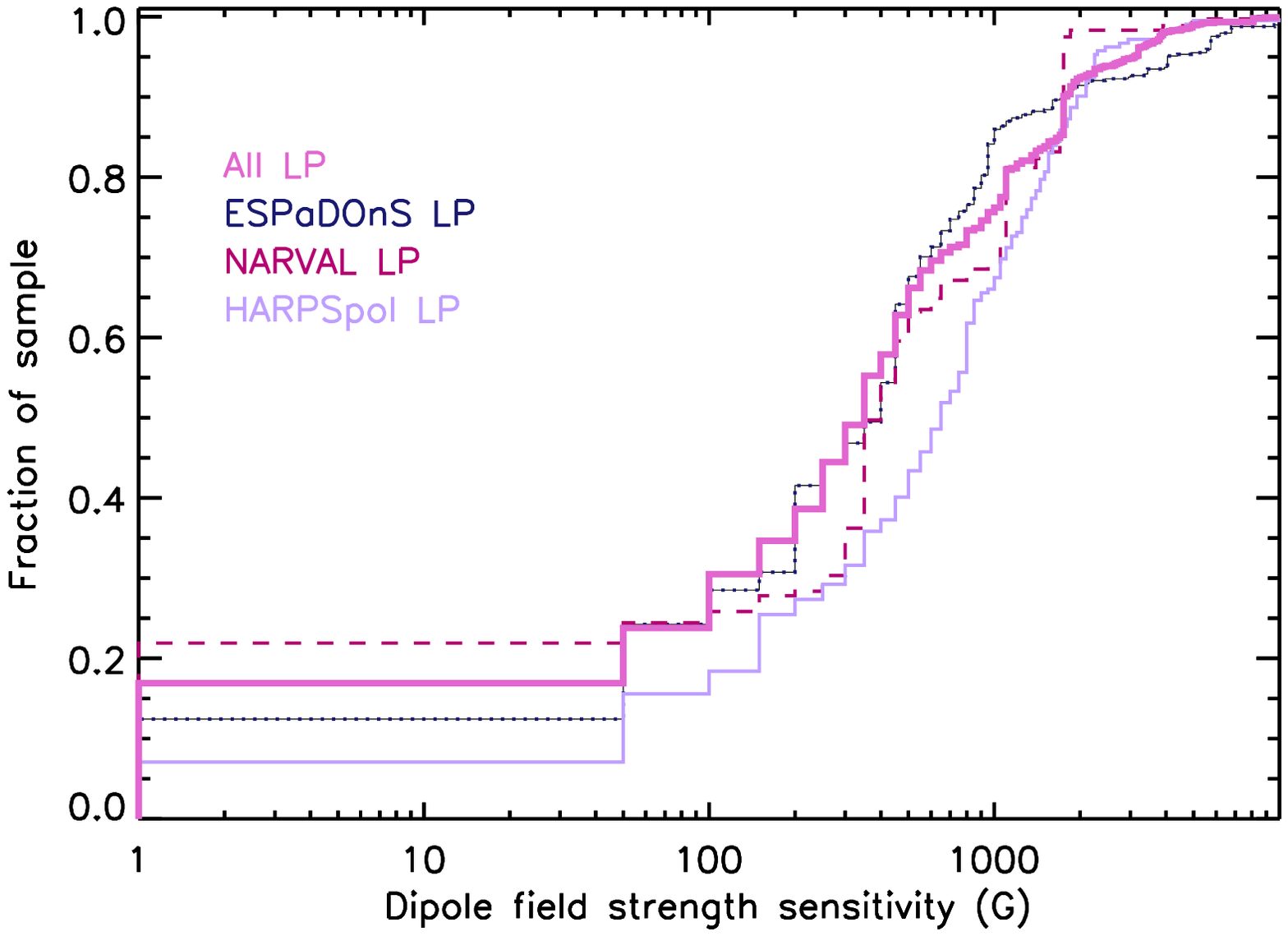}\hspace{1cm}\includegraphics[width=8cm]{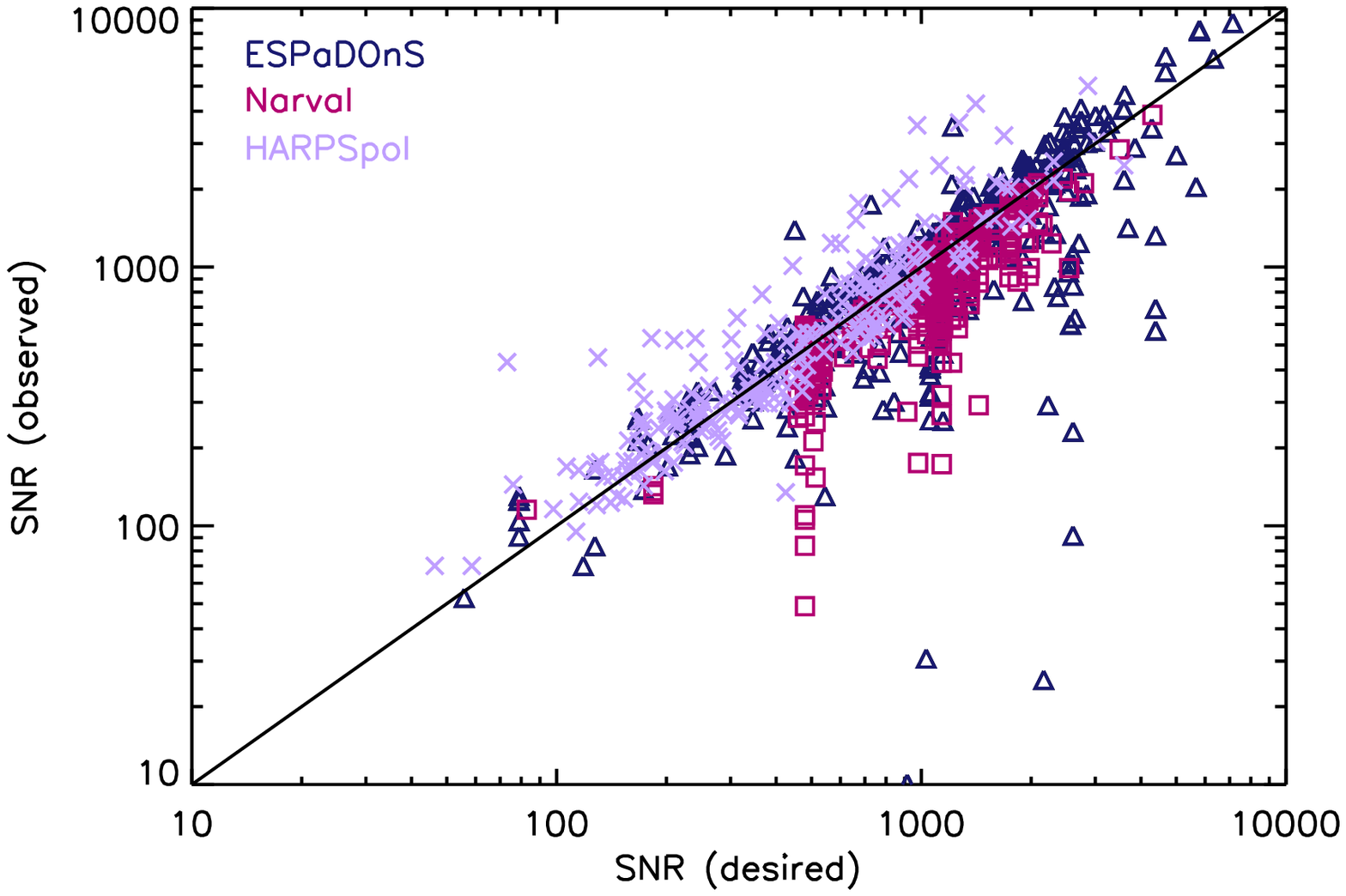}
\caption{\label{expresults}{\em Left -}\ Cumulative histograms of the predicted dipole magnetic field strength sensitivity, according to the SNRs achieved during LP observations.  {\em Right -}\ Achieved SNRs versus those predicted according to exposure time for all SC LP observations, according to instrument (triangles for ESPaDOnS, squares for Narval and crossed for HARPSpol).}
\end{centering}
\end{figure*}

For each star in a given sensitivity bin, the exposure time was adjusted to achieve a SNR$_{\rm LSD}$ following application of LSD that allowed the detection of that field strength. For practical purposes, targets were typically assigned to the most sensitive bin for which the required exposure time for that star was below about 2 hours.  Consequently, for some targets nominally assigned to the 1 kG bin the required SNR was not achievable within this practical time limit. As a result, about 25\% of the LP observations (corresponding to about 90 targets) yield predicted dipole field strength sensitivities that are larger than 1~kG (Fig.~\ref{expresults}, left frame).

The approximate relations governing the spectrum SNR required to reach a magnetic precision $B_{\rm 0.1}$ in units of 0.1~kG were determined through empirical fits to the model results:

\begin{equation}
\label{eqntime1}
{\rm SNR_{\rm LSD}} = (120+170\times v\sin i)\, B_{\rm 0.1}^{-1}\ \  [{\rm if}\ v\sin i\leq 40~{\rm km/s}];
\end{equation}

\noindent or

\begin{equation}
\label{eqntime2}
{\rm SNR_{\rm LSD}} = (-18700+640\times v\sin i)\, B_{\rm 0.1}^{-1}\ \  [{\rm if}\ v\sin i> 40~{\rm km/s}].
\end{equation}

The accuracy of these empirical relations will be evaluated in forthcoming papers.



The total exposure time (in seconds) required was then computed by first dividing the required LSD SNR by the inferred LSD gain factor $G({\rm ST})$ to obtain the required SNR in the reduced spectrum, SNR$_{\rm spec}$. Finally, we applied the appropriate official exposure time relation ETC$(V,{\rm SNR_{spec}})$ for each instrument to infer the exposure time\footnote{With the replacement of the ESPaDOnS EEV1 chip with Olapa in 2010, the ESPaDOnS exposure time calculator (ETC) was updated to reflect the new detector characteristics. MiMeS exposure times were also updated to compensate.}\footnote{During the first HARPSpol observing runs, it was identified that the exposure time predictions of the HARPSpol ETC strongly overestimated the actual SNRs achieved. Therefore, in subsequent runs, exposure times were increased by a factor of 2.25, leading to an increase of 50\% in SNR.}.

In many cases, the required aggregate spectrum SNR was too high to be achieved in a single observation without saturation. In these cases, the observation was subdivided into several subsequences. The total time required to obtain an observation of many hot, bright and/or broad-lined stars was therefore often dominated by overheads.

For example, for HD~87901 (Regulus, B8IVn, $V=1.4$, $v\sin i\simeq 300$~km/s) the SNR$_{\rm LSD}$ required for a magnetic sensitivity of 250~G (i.e. $B_{\rm 0.1}=2.5$) was about 70,000. For a gain factor consistent with its spectral type ($G({\rm B8V})=14$), the required SNR$_{\rm spec}$ in the aggregate spectrum was computed to be about 5,000. Observations were acquired with ESPaDOnS. The ESPaDOnS ETC predicted a maximum exposure time (before saturation) per polarimetric subexposure of 10 s. Sixteen observations corresponding to 4 subexposures of 10 s each were acquired. The total exposure time was 640 s, whereas the total observing time including official overheads was 3200 s. Hence the overheads corresponded to 80\% of the total observing time required.

For the actual observations of HD~87901, the combined SNR in the coadded Stokes $V$ spectrum was 5700, leading to an expected magnetic sensitivity (based on Eq.~\ref{eqntime2} and the observed SNR) of about 220~G.


In addition to the LP observations, a significant fraction of the SC observations were collected from the archives. Hence the exposure times and sensitivities of these observations are diverse, and adopted by the original PIs according to their scientific goals.

\subsection{Quantitative magnetic diagnosis}

The quantitative determination of the detection of a magnetic signature (e.g. Fig.~\ref{hd175362}) in the LSD profile is obtained in two ways. First, we use the Stokes $V$ spectra to measure the mean longitudinal magnetic field strength \bz\ of each star at the time of observation. We can also examine spectral lines for the presence of circular polarisation signatures: Zeeman splitting combined with Doppler broadening of lines by rotation leads to non-zero values of $V$ within spectral lines even when the value of \bz\ is equal to zero. This possibility substantially increases the sensitivity of our measurements as a discriminant of whether a star is in fact a magnetic star or not, as discussed by \citet{2002A&A...392..637S, 2009MNRAS.398.1505S} and \citet{2012ApJ...750....2S}.


The field \bz\ is obtained by integrating the $I/I_{\rm c}$ and $V/I_{\rm c}$ profiles (normalized to the continuum $I_{\rm c}$) about their centres-of-gravity $v_{\rm 0}$ in velocity $v$, in the manner implemented by \citet{1979A&A....74....1R,1997MNRAS.291..658D} and corrected by \citet{2000MNRAS.313..851W}:

\begin{equation}
\label{eqnbz}
\bz=-2.14\times 10^{11}\ \frac{{\displaystyle \int (v-v_{\rm 0}) V(v)\ dv}}{\displaystyle {\lambda z c\ \int [1-I(v)]\ dv}}.
\end{equation}

In Eq.~(\ref{eqnbz}), $V(v)$ and $I(v)$ are the $V/I_{\rm c}$ and $I/I_{\rm c}$ profiles, respectively. The wavelength $\lambda$ is expressed in nm and the longitudinal field \bz\ is in gauss. The wavelength and Land\'e factor $z$ correspond to those used to normalize the LSD profile at the time of extraction. Atomic data were obtained from the Vienna Atomic Line Database (VALD) where available. When experimental Land\'e factors were unavailable, they were calculated assuming L-S coupling. The limits of integration are usually chosen for each star to coincide with the observed limits of the LSD $I$ and $V$ profiles ; using a smaller window would neglect some of the signal coming from the limb of the star, while a window larger than the actual line would increase the noise without adding any further signal, thus degrading the SNR below the optimum value achievable \citep[see e.g.][]{2012A&A...546A..44N}.

In addition, the LSD Stokes $V$ profile is itself examined. We evaluate the false alarm probability (FAP) of $V/I_{\rm c}$ inside the line according to:

\begin{equation}
{\rm FAP}(\chi^2_r,\nu)=1-P({\nu\over {2}},{{\nu\chi^2_r}\over{2}}),
\end{equation}

\noindent where $P$ is the incomplete gamma function, $\nu$ is the number of spectral points inside the line, and $\chi^2_r$ is the reduced chi-square ($\chi^2/\nu$) computed across the $V$ profile  \citep[e.g.][]{1992A&A...265..669D}. The  reference level required to compute $\chi^2/\nu$, while in principle equal to $V=0$, may be affected by small offsets related to instrumentation and data reduction. In this work, to avoid potential systematics related to such offsets, we employ the mean of $V$, measured outside of the spectral line, as the reference for calculation of $\chi^2/\nu$. The FAP value gives the probability that the observed $V$ signal inside the spectral line could be produced by chance if there is actually no field present. Thus a very small value of the FAP implies that a field is actually present. We evaluate FAP using the detection thresholds of \citet{1997MNRAS.291..658D}. We consider that an observation displays a ``definite detection" (DD) of Stokes $V$ Zeeman signature if the FAP is lower than 0.00001, a ``marginal detection" (MD) if it falls between 0.001 and 0.00001, and a ``null detection" (ND) otherwise. As mentioned above a significant signal (i.e. with a MD or DD) may occur even if \bz\ is not significantly different from zero. Normally, a star was considered to have been detected if a significant signal (i.e. with a MD or DD) was detected within the line, while always remaining insignificant in the neighbouring continuum and in the $N$ profile.





\section{Polarimetric performance and quality control}
\label{performance}

\subsection{Overview of data quality}

Data quality was quantified and monitored in several ways during acquisition and analysis. 

We adopt SNR per spectral pixel in the reduced, 1-dimensional polarimetric spectra as our principal indicator of data quality. For ESPaDOnS and Narval spectra, this corresponds to a 1.8~km/s pixel measured in the null spectrum, whereas for HARPSpol spectra, the spectral bin is 0.8~km/s. SNR is defined as the inverse of the formal uncertainty of each pixel normalized to the continuum, and is determined from counting statistics by tracking photons through the entire spectral reduction process. As a consequence, each reduced spectrum is accompanied by error bars (i.e. 1/SNR) associated with each pixel. The accuracy of the SNR calculation is verified using measurements of the RMS deviation in the diagnostic null. 

\begin{figure}
\begin{centering}
\includegraphics[width=8cm]{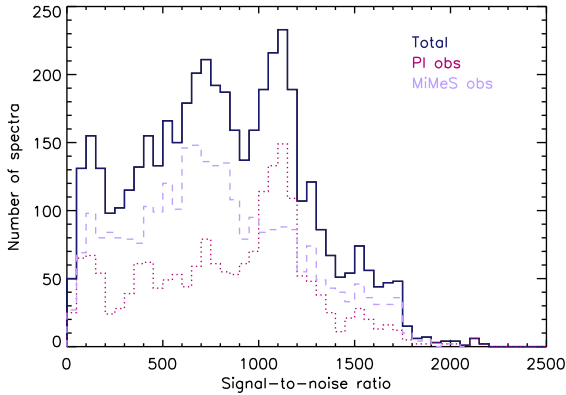}
\caption{\label{snrs}Distributions of SNRs per spectral pixel at 500 nm. For comparison, the spectrum shown in Fig. 1 has a SNR of 860. Different colours and linestyles are used to distinguish between all, LP, and archival (PI) observations.}
\end{centering}
\end{figure}

The distribution of SNRs of the TC and SC spectra is illustrated in Fig.~\ref{snrs}. The distribution is very broad, extending from values of a few tens, and with a tail extending to $>2000$. The median SNR is 800. The breadth and structure of the distribution can be ascribed to three factors. First, recall that the desired SNR of each SC target was computed in order to achieve a particular magnetic sensitivity,  and that such a calculation is a function of the stellar spectral characteristics (spectral type, $v\sin i$; see Fig.~\ref{exptimemodel}). Hence stars with different spectral characteristics can require significantly different SNRs to achieve the same magnetic sensitivity. Moreover, as described in \S\ref{magsens}, a range of magnetic sensitivity targets was adopted in this study. Secondly, recall that TC targets were observed repeatedly, and that the observations of a particular TC target typically have roughly the same SNR. Finally, archival data included in the SC have diverse SNR characteristics that were presumably determined by the scientific requirements of the associated programs. The form and structure of the SNR distribution are mainly a consequence of these effects, in addition to poor weather.

Fig. \ref{expresults} (left panel) shows the cumulative histograms of the predicted surface dipole field strength sensitivities, based on the SNRs achieved during the LPs. For the combined sample, 50\% of observations are estimated to be sensitive to surface dipole magnetic fields equal to or stronger than 375~G. Note that, in particular, for 75\% of the observed sample we predict sensitivity to dipole fields of 1 kG or weaker. These predicted sensitivities will be evaluated in greater detail in future papers. The right panel summarizes the achieved SNRs per spectral pixel as compared to the desired SNRs computed using the exposure time model, for the LP SC observations. The results are in reasonable agreement with the 1:1 relationship, indicating that the dataset fulfils the initial requirements. 



As is discussed by \citet{2012MNRAS.426.1003S}, the ESPaDOnS and Narval instruments exhibit small differences in resolving power (2-3\%) relative to each other, and small variations of resolving power with time. Such small differences and variations should have no significant impact on the quality of the magnetic measurements. Our data are consistent with these conclusions. \citet{2012MNRAS.426.1003S} also demonstrate the good agreement between magnetic analyses performed using ESPaDOnS and Narval. 

The HARPSpol instrument differs from ESPaDOnS and Narval in terms of its general design and optical strategy, ultimately leading to polarized spectra covering a smaller wavelength window but with significantly higher resolution. Due to the locations of the instruments in different hemispheres, there are as yet few examples of magnetic stars that have been monitored by both HARPSpol and the northern instruments in order to verify their spectral and polarimetric agreement in detail. However, \citet{2011Msngr.143....7P} (in their figure 5) illustrate the agreement of the Stokes $I$ and $V$ spectra of the sharp-lined Ap star $\gamma$~Equ, and \citet{2015A&A...575A.115B} demonstrate (in their figure 1) that the longitudinal field of HD 94660 as measured by ESPaDOnS agrees with the variation inferred form HARPSpol measurements.

\subsection{TC targets as magnetic and spectral standards}
\label{SectTC}

The principal method of monitoring the accuracy and precision of the polarimetric analysis of all 3 instruments was through the examination of the recurrent observations of magnetic stars (typically TC targets). 

\begin{figure}
\begin{centering}
\subfloat[][Longitudinal field versus HJD]{\includegraphics[width=8cm]{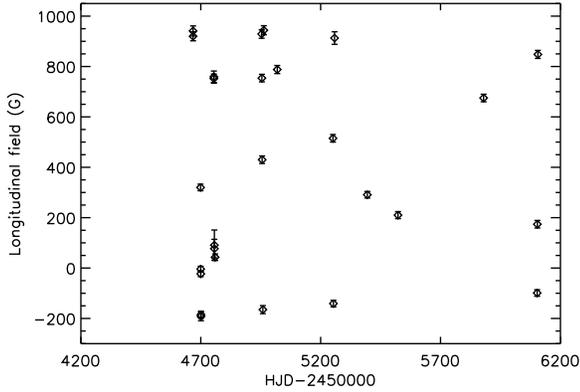}}

\subfloat[][Longitudinal field versus phase]{\includegraphics[width=8cm]{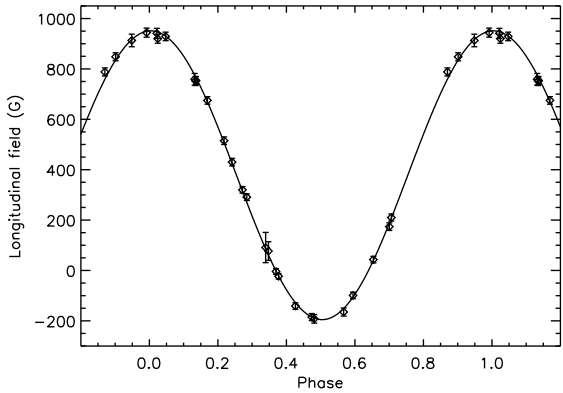}}

\subfloat[][Null field versus phase]{\includegraphics[width=8cm]{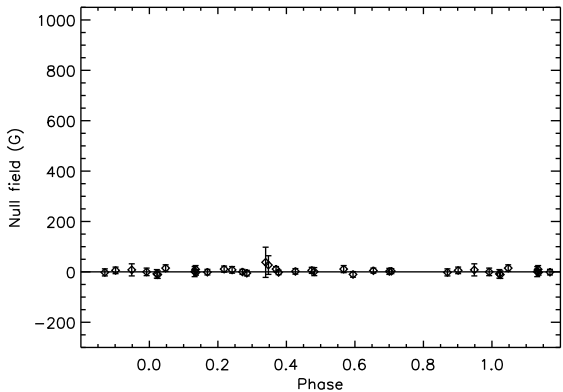}}
\caption{\label{hd184927}Longitudinal field measurements of the strong-field ESPaDOnS TC target HD 184927 ($P_{\rm rot}=9.53$~d). Adapted from \citet{2015MNRAS.447.1418Y}.}
\end{centering}
\end{figure}

\begin{figure}
\begin{centering}
\subfloat[][Longitudinal field versus HJD]{\includegraphics[width=8cm]{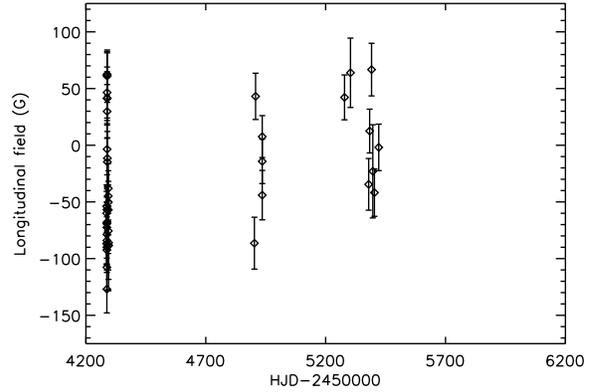}}

\subfloat[][Longitudinal field versus phase]{\includegraphics[width=8cm]{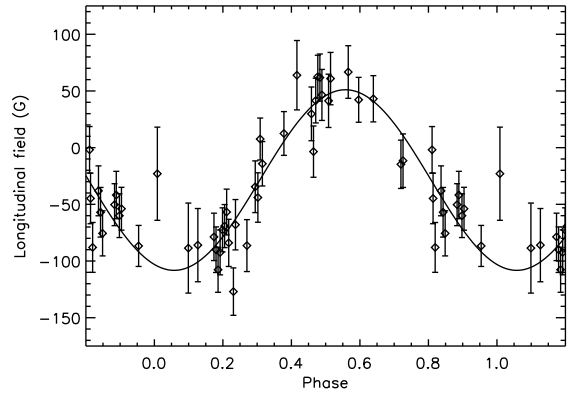}}

\subfloat[][Null field versus phase]{\includegraphics[width=8cm]{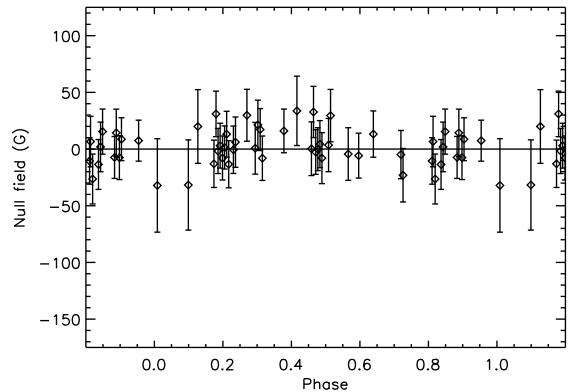}}
\caption{\label{v2052oph}Longitudinal field measurements of the weak-field Narval TC target V2052 Oph ($P_{\rm rot}=3.64$~d). Adapted from \citet{2012A&A...537A.148N}.}
\end{centering}
\end{figure}

Repeated observations of many TC targets confirm their strict periodicity on the timescale of the MiMeS observations \citep[e.g.][]{2011MNRAS.416.3160W,2012MNRAS.426.2208G,2015MNRAS.447.1418Y}. This periodic variability, on timescales ranging from less than 1 day to more than 1 year, provides a powerful method to verify the long-term stability of the polarimetric performance of the instruments, as well as the compatibility of their magnetic analyses. Figures~\ref{hd184927} and \ref{v2052oph} illustrate the longitudinal magnetic field variations, from both the Stokes $V$ and diagnostic $N$ profiles (shown at the same display scale as $V$), for two MiMeS TC targets: HD 184927, a strong-field early Bp star studied by \citet{2015MNRAS.447.1418Y}, and V2052 Oph, a weak-field $\beta$~Cep star studied by \citet{2012A&A...537A.148N}.

For HD 184927, 28 good-quality Stokes $V$ measurements were obtained with ESPaDOnS between HJD 2454667 (July 20 2008) and 2456105 (June 27 2012), corresponding to 1438 days or approximately 4 years of observation. The rotational period of HD 184927 is 9.53 days, and the time over which the data were acquired corresponds to more than 150 stellar rotations. The median error bar of the longitudinal field measurements from LSD profiles is 15 G, and the reduced $\chi^2$ of a sinusoidal fit with fixed period is 0.6. Clearly all of the measurements of HD 184927 agree very well with a sinusoidal variation stable within $\sim15$~G during the period 2008-2012.

For V2052 Oph, 44 good-quality Stokes $V$ measurements were obtained with Narval between HJD 2454286 (July 4 2007) and HJD 2455421 (August 12 2010), corresponding to 1135 days or approximately 3.1 years of observation. The rotational period of V2052 Oph is 3.64 days, and the time over which the data were acquired corresponds to more than 300 stellar rotations. The median error bar of the longitudinal field measurements from LSD profiles is 21 G. A purely sinusoidal fit provides a good reproduction of the phase variation of the observations, resulting in a reduced $\chi^2$ of 1.2. These results are consistent with those reported by \citet{2012A&A...537A.148N}, and demonstrate the long-term repeatability of measurements of even a relatively weak magnetic field. All of the measurements of V2052 Oph agree well with this unique harmonic variation stable within $\sim20$~G during the period 2007-2010.

The long-term agreement of these measurements provides confidence that no unidentified instrumental changes (e.g. associated with instrument mounting/dismounting, change of the ESPaDOnS CCD, short-term and long-term drifts, etc.) have occurred during the MiMeS project. It also demonstrates that the measurements are insensitive to the ESPaDOnS instrumental crosstalk, which was systematically reduced from $\sim5$\% to below 1\% during the course of the project.

In the context of the recent examinations of magnetometry obtained with the low-resolution FORS spectropolarimeters \citep{2012A&A...538A.129B,2014A&A...572A.113L}, Figs.~\ref{hd184927} and \ref{v2052oph} are of great interest. In contrast to FORS1, there does not, except for the short period of malfunction of the Narval rhomb \#2, discussed in \S\ref{SectNarval}), seem to be any problem of occasional statistically significant outliers. As a result, small data sets can be safely used to estimate periods, for example, without fear that the period obtained is badly polluted by one 4-5$\sigma$ outlier. 

\begin{figure*}
\begin{centering}
\hspace{-0.4cm}\includegraphics[width=6cm]{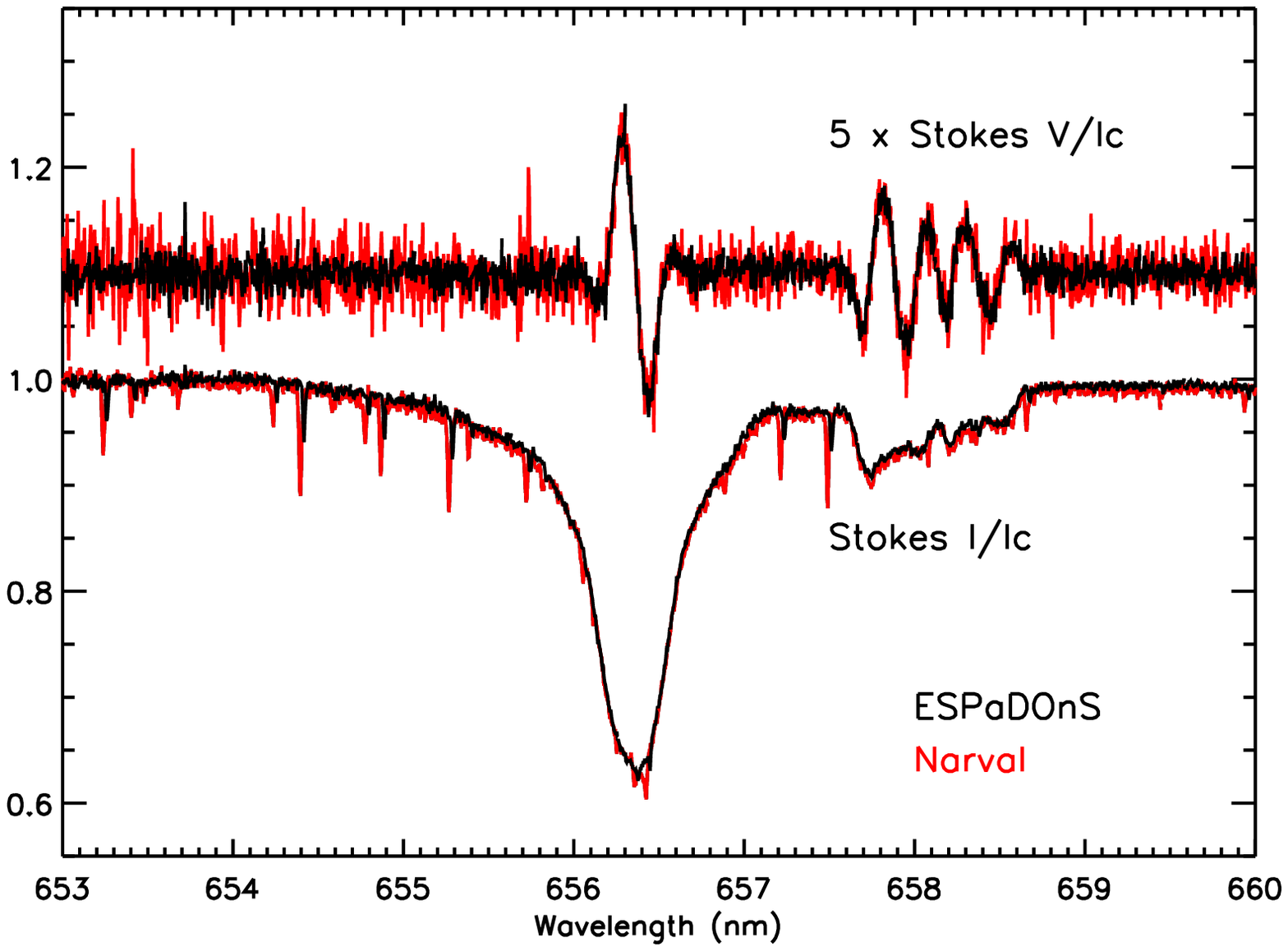}\includegraphics[width=6cm]{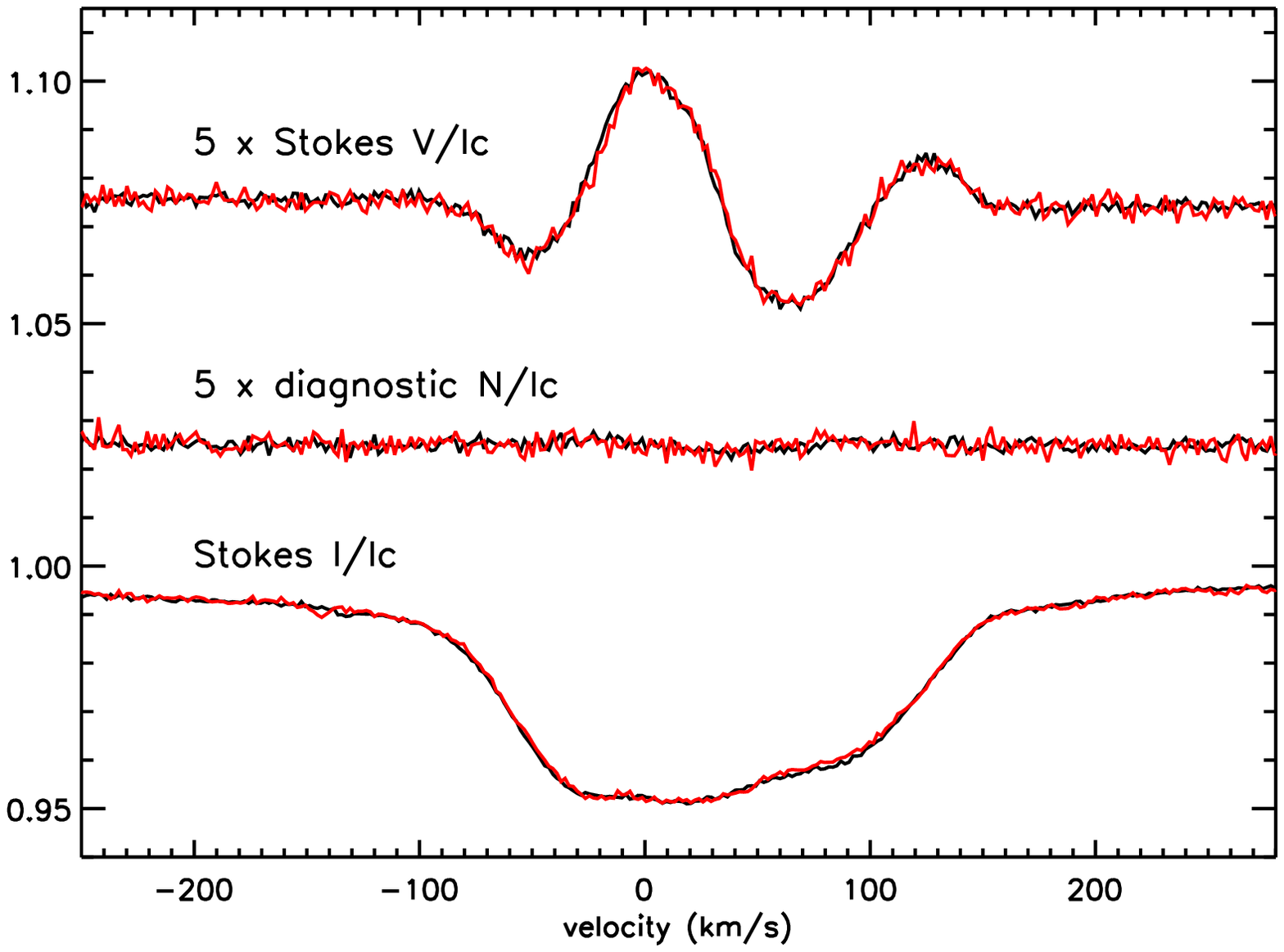}\includegraphics[width=6cm]{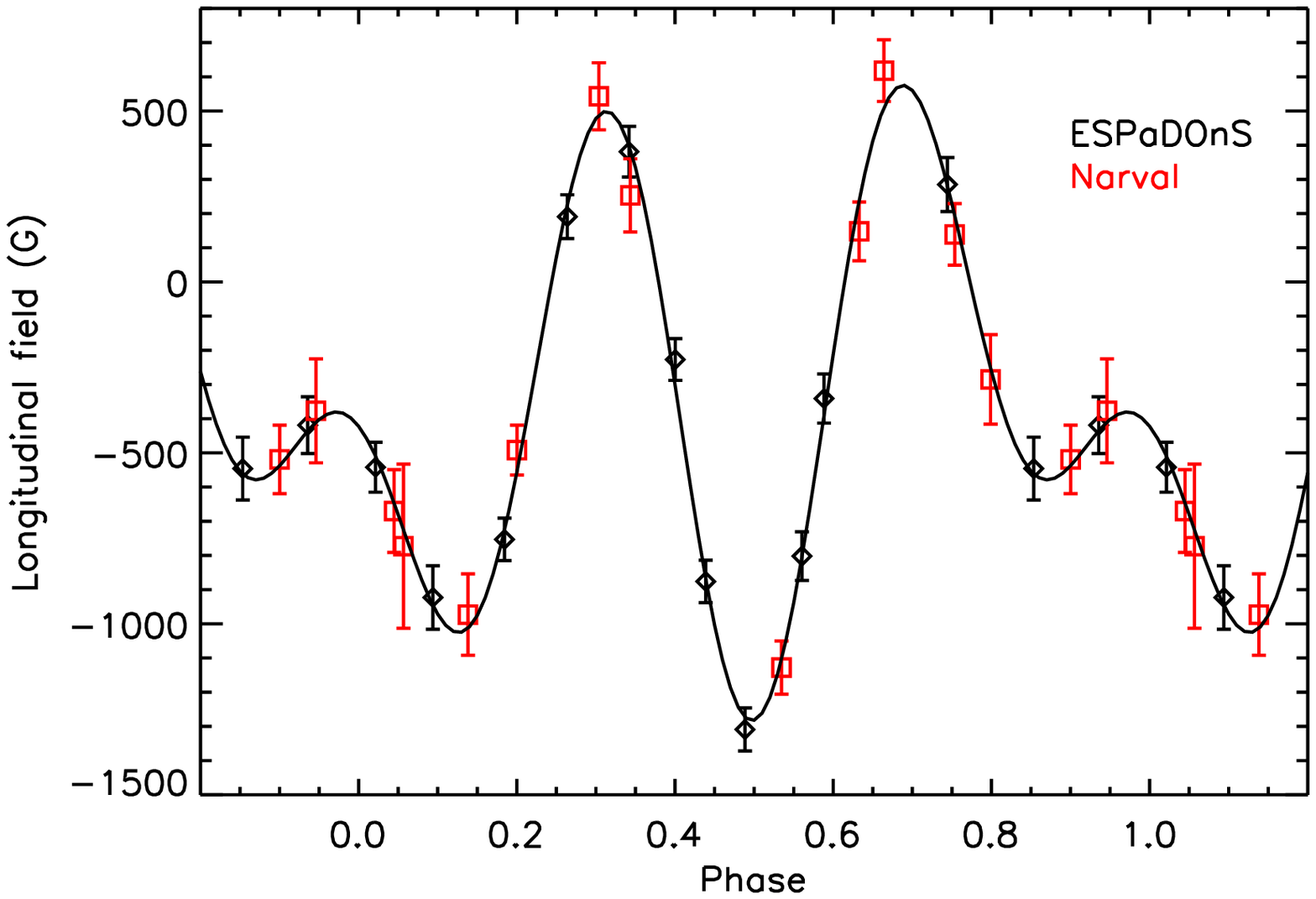}
\caption{\label{hd37776}Comparison of ESPaDOnS and Narval observations of HD 37776 (rotational period $P_{\rm rot}=1.54$~d). {\em Left -}\ Comparison of ESPaDOnS (black) and Narval (red) Stokes $I$ and $V$ spectra of the TC target HD 37776 at phase 0.34, obtained 21 days apart. A small part of the red region of the spectrum showing the H$\alpha$ line. {\em Middle -}\ Comparison of LSD profiles extracted from the full ESPaDOnS and Narval spectra at the same phase. {\em Right -}\ Comparison of all longitudinal field measurements of HD 37776 obtained with ESPaDOnS and Narval, phased according to the ephemeris of \citet{2008A&A...485..585M}. The solid curve is a 3rd-order harmonic fit to the combined data. Observations were obtained between JDs 2454845 and 2455967, i.e. over a period of more than 3 years. The internal and external agreement of the datasets is excellent. Adapted from Shultz et al., in prep.}
\end{centering}
\end{figure*}

Secondly, another problem identified clearly with FORS data is the need to ensure that all measurements are on the same instrumental system. This is especially important when constructing magnetic curves and using them to determine new or improved periods. It has been established for FORS1 \citep{2014A&A...572A.113L} that each choice of grism and wavelength window constitutes a distinct instrumental measuring system, and that simultaneous field measurements in different instrumental systems may result in significantly different field strengths. Fig.~\ref{hd37776} shows spectra and LSD profiles of the magnetic TC target HD 37776 acquired using two different instruments (ESPaDOnS and Narval), at the same rotational phase on dates separated by about 21 days. The Stokes $I$ and $V$ profiles are identical to within the uncertainties. The right panel shows phased longitudinal field measurements obtained with both ESPaDOnS and Narval for the same star, demonstrating that the two instrumental systems are essentially identical and data from the two instruments may be confidently combined. This includes \bz\ measurements and LSD profiles (e.g. Fig.~\ref{hd37776}), as long as they are extracted using the same line mask applied to the same spectral regions (which was the case for all MiMeS observations).

Examples of MiMeS observations of TC targets acquired with the HARPSpol instrument are reported by \citet{2011A&A...536L...6A,2014A&A...567A..28A}. These observations span a shorter time than those described above. Other monitoring observations \citep[such as those of][]{2013A&A...558A...8R} better demonstrate the long-term stability of the HARPSpol instrument.  

These examples, published and proprietary observations of other MiMeS TC targets (see Table~\ref{TC}), and complementary published results \citep[e.g.][]{2012MNRAS.426.1003S} provide a strong verification of the long term stability of the sensitivity, zero point, and scale of magnetic measurements acquired with ESPaDOnS and Narval. HARPSpol was commissioned only in 2011, so more limited data exist with which to evaluate its long-term stability and compatibility of its measurements with those obtained with other spectropolarimeters. However, the existing data suggest very good agreement and stability.


\section{Summary and conclusions}
\label{summary}

The MiMeS survey of Magnetism in Massive Stars is by far the largest systematic
survey of massive star magnetism ever undertaken. The goal of this project is to
unravel the origin of magnetic fields in massive stars, and to understand the impact of
magnetic fields on stellar mass loss and environment, rotational evolution, and ultimately stellar
evolution.

This paper has described the methodology of the project. Many papers reporting analyses of TC targets have already been
published in conference proceedings and refereed journals \citep[e.g.][]{2010MNRAS.405L..51O, 2011MNRAS.416.3160W,2012A&A...537A.148N,2012MNRAS.423..328B,2014A&A...565A..83K,2015MNRAS.447.1418Y}. In addition, many of
the magnetic stars that were newly-detected or confirmed during the SC have been followed
up and are discussed in refereed papers. For the O stars, these include
\citet{2009MNRAS.400L..94G,2012MNRAS.426.2208G,2013MNRAS.428.1686G,2010MNRAS.407.1423M,2012MNRAS.419.2459W,2012MNRAS.425.1278W,2015MNRAS.447.2551W} and 
\citet{2013MNRAS.433.2497S}. For B stars, these
include \citet{2011MNRAS.412L..45P,2011A&A...536L...6A,2012MNRAS.419.1610G,2013A&A...557L..16B,2014A&A...563L...7N,2014A&A...567A..28A}, and \citet{2015MNRAS.449.3945S}.


Similarly, some TC and SC null results of particular significance have already
been published. These include measurements of Wolf-Rayet stars
\citep{2013ApJ...764..171D, 2014ApJ...781...73D}, bright O and B stars
exhibiting DACs \citep{2014MNRAS.444..429D}, BA supergiants
\citep{2014MNRAS.438.1114S}, and a number of B stars in which detections of
magnetic fields were previously claimed but that were not confirmed by
independent MiMeS observations \citep{2012ApJ...750....2S}.

The present paper has concentrated on the SC. The survey comprises over 4800
circularly polarized spectra of 106 O and WR stars, and 422 B stars, ranging from spectral
type O4 to B9.5 $V\sim 0$ to 13.6. We have acquired data of
these 528 stars thanks to large programs of observations with the three
high resolution spectropolarimeters available in the world: ESPaDOnS@CFHT,
Narval@TBL, and HARPSpol@ESO. We have established the reliability of the
observational tools by comparing the data obtained from the three instruments,
as well as the obtained versus initially expected quality of the data. We have shown
that the data are mutually consistent and perfectly suitable for our science goals. 

In particular, these high resolution, high SNR spectropolarimetric data allow
us to determine the fundamental parameters of each target \citep[see][for the O
stars]{2015A&A...575A..34M}, as well as the magnetic field and magnetospheric
properties \citep[e.g.][]{2013MNRAS.429..398P}. While this paper introduces the MiMeS
survey, a series of forthcoming papers will present the magnetic analysis of
several subsamples of stars: the O, B, classical Be, pulsating OB, OB
supergiants, and cluster stars. Interpretation of the null results for all O and
B stars in terms of upper field limits will also be published, as well as the
fundamental parameters of the B stars \citep[for the O stars,
see][]{2015A&A...575A..34M}. Ultimately, the survey results will allow us to
quantify the occurence of magnetic fields in massive stars and search for
correlation between the properties of magnetic fields and stellar properties.

This article represents the introduction to the MiMeS survey. Nine additional papers related to the SC are currently planned or in preparation:

\begin{itemize}
\item Magnetic analysis of the O-stars sample (Grunhut et al., in prep)
\item Interpretation of the O-stars null results (Petit et al., in prep)
\item Magnetic analysis of the classical Be stars sample (Neiner et al., in prep)
\item Magnetic analysis of the O and B supergiants sample (Oksala et al., in prep)
\item Magnetic analysis of the open clusters sample (Alecian et al., in prep)
\item Magnetic analysis of the pulsating OB stars (Neiner et al., in prep)
\item Magnetic analysis of the B-stars sample (Grunhut et al., in prep)
\item Interpretation of the B-stars null results (Petit et al., in prep)
\item Physical parameters of the B-stars sample (Landstreet et al., in prep)
\end{itemize}

 \section*{Acknowledgments}
CFHT, TBL and HARPSpol observations were acquired thanks to generous allocations of observing time within the context of the MiMeS Large Programs. EA, CN, and the MiMeS collaboration acknowledge financial support from the Programme National de Physique Stellaire (PNPS) of INSU/CNRS. This research has made extensive use of the {\sc simbad} database, operated at CDS, Strasbourg, France. We acknowledge the Canadian Astronomy Data Centre (CADC). GAW, AFJM and JDL acknowledge support from the Natural Science and Engineering Research Council of Canada (NSERC). AFJM acknowledges support from the Fonds de Recherche de Qu\'ebec - Nature et Technologies (FRQNT). YN acknowledges support from  the Fonds National de la Recherche Scientifique (Belgium), the PRODEX XMM contracts, and the ARC (Concerted Research Action) grant financed by the Federation Wallonia-Brussels. OK is a Royal Swedish Academy of Sciences Research Fellow, supported by the grants from the Knut and Alice Wallenberg Foundation, the Swedish Research Council, and the G\"oran Gustafsson Foundation. SPO acknowledges partial support from NASA Astrophysics Theory Program grant NNX11AC40G to the University of Delaware. RHDT, AuD and SPO acknowledge support from NASA grant NNX12AC72G. JOS acknowledges funding from the European Union's Horizon 2020 research and innovation programme under the Marie Sklodowska-Curie grant agreement No 656725. CPF acknowledges support from the French ANR grant {\em Toupies: Towards understanding the spin evolution of stars.} TL acknowledges funding of the Austrian FFG within ASAP11 and the FWF NFN project S116601-N16.  AuD and DHC acknowledge support by NASA through Chandra Award numbers TM4-15001A, TM4-15001B issued by the Chandra X-ray Observatory Center which is operated by the Smithsonian Astrophysical Observatory for and behalf of NASA under contract NAS8-03060.  RHB acknowledges support from FONDECYT Project No 1140076. WWW was supported by the Austrian Science Fund (FWF P22691-N16). IY acknowledges support from Russian Scientific Foundation (RSF grant number 14-50-0043).We thank Greg Barrick and Pascal Petit for their assistance with technical details of the instrumentation. The construction of ESPaDOnS was supported in part by a Major Installation grant from NSERC. The authors extend their warm thanks to the staff of the CFHT and TBL for their efforts in support of the MiMeS project.

\bibliography{survey1}

\clearpage

\begin{table*}																																																
\caption{\label{SCO}MiMeS O-type and WR SC targets. Columns report common identifier and HD \#, $V$ band magnitude, the number of ESPaDOnS (E), Narval (N) and HARPSpol (H) spectra acquired, and the total number of spectra, the spectral type, luminosity class (LC) and any spectral peculiarity (pec), and the reference to the spectral type, luminosity class and peculiarity.}																						
\begin{center}\begin{tabular}{llllllllllllllrrc}																																																
\hline																																							& & & \multicolumn{4}{c}{\# of observations}\\									
Name		&		HD		&		V			&		E		&		N		&		H		&		Tot			&		Spectral type		&		LC		&		pec		&		Reference	\\	
\hline																																																
 HD 108       		&		108		&		7.38			&		23		&		87		&		0		&		110			&		 O8    		&		         		&		 f?p var      		&		 \citet{2011ApJS..193...24S}  	\\	
 AO Cas       		&		1337		&		6.1	 		&		1		&		0		&		0		&		1			&		 O9.5  		&		 II      		&		 (n)         		&		 \citet{2011ApJS..193...24S}  	\\	
WR 1		&		4004		&		10.14	 		&		4		&		0		&		0		&		4			&		WN4b		&				&				&		\citet{crowther}	\\	
 V354 Per     		&		13745		&		7.9	 		&		0		&		3		&		0		&		3			&		 O9.7  		&		 II      		&		 (n)         		&		 \citet{2011ApJS..193...24S}  	\\	
 HD 14633     		&		14633		&		7.47	 		&		1		&		0		&		0		&		1			&		 ON8.5 		&		 V       		&		             		&		 \citet{2011ApJS..193...24S}  	\\	
 HD 17505     		&		17505		&		7.1	 		&		1		&		0		&		0		&		1			&		 O6.5  		&		 III     		&		 n((f))      		&		 \citet{2011ApJS..193...24S}  	\\	
 HD 24431     		&		24431		&		6.8	 		&		2		&		0		&		0		&		2			&		 O9    		&		 III     		&		             		&		 \citet{2011ApJS..193...24S}  	\\	
 X Per        		&		24534		&		6.72	 		&		2		&		0		&		0		&		2			&		 O9.5: 		&		         		&		 npe         		&		 \citet{2011ApJS..193...24S}  	\\	
 $\xi$ Per      		&		24912		&		4.06	 		&		5		&		39		&		0		&		44			&		 O7.5  		&		 III     		&		 (n)((f))    		&		 \citet{2011ApJS..193...24S}  	\\	
 $\alpha$ Cam      		&		30614		&		4.3	 		&		10		&		1		&		0		&		11			&		 O9    		&		 Ia      		&		             		&		 \citet{2011ApJS..193...24S}  	\\	
 AE Aur       		&		34078		&		6	 		&		1		&		11		&		0		&		12			&		 O9.5  		&		 V       		&		             		&		 \citet{2011ApJS..193...24S}  	\\	
 HD 34656     		&		34656		&		6.8	 		&		1		&		0		&		0		&		1			&		 O7.5  		&		 II      		&		 (f)         		&		 \citet{2011ApJS..193...24S}  	\\	
 HD 35619     		&		35619		&		8.66	 		&		1		&		0		&		0		&		1			&		 O7.5  		&		 V       		&		 ((f))       		&		 \citet{2011ApJS..193...24S}  	\\	
 LY Aur       		&		35921		&		6.85	 		&		1		&		0		&		0		&		1			&		 O9.5  		&		 II      		&		             		&		 \citet{2011ApJS..193...24S}  	\\	
$\delta$ Ori A		&		36486		&		2.41			&		0		&		11		&		0		&		11			&		O9.5		&		II		&		Nwk		&		\citet{2011ApJS..193...24S}	\\	
$\upsilon$ Ori		&		36512		&		4.62	 		&		0		&		1		&		0		&		1			&		O9.7		&		V		&				&		\citet{2011ApJS..193...24S}	\\	
 $\lambda$ Ori A    		&		36861		&		3.3	 		&		9		&		11		&		0		&		20			&		 O8    		&		 III     		&		 ((f))       		&		 \citet{2011ApJS..193...24S}  	\\	
 HD 36879     		&		36879		&		7.58	 		&		3		&		0		&		0		&		3			&		 O7    		&		 V       		&		 (n)((f))    		&		 \citet{2011ApJS..193...24S}  	\\	
 43 Ori                   		&		37041		&		5.08	 		&		11		&		0		&		0		&		11			&		   O9.5		&		IV 		&		p  		&		 \citet{2011ApJS..193...24S} 	\\	
 $\iota$ Ori      		&		37043		&		2.77	 		&		12		&		0		&		0		&		12			&		 O9    		&		 III     		&		 var         		&		 \citet{2011ApJS..193...24S}  	\\	
 HD 37366     		&		37366		&		7.64	 		&		1		&		0		&		0		&		1			&		 O9.5  		&		 IV      		&		             		&		 \citet{2011ApJS..193...24S}  	\\	
 $\sigma$ Ori      		&		37468		&		3.8	 		&		5		&		0		&		0		&		5			&		 O9.7  		&		 III     		&		             		&		 \citet{2011ApJS..193...24S}  	\\	
 $\mu$ Col       		&		38666		&		5.15	 		&		2		&		0		&		0		&		2			&		 O9.5  		&		 V       		&		             		&		\citet{2013msao.confE.101S}	\\	
 HD 42088     		&		42088		&		7.56	 		&		1		&		0		&		0		&		1			&		 O6    		&		 V       		&		 ((f))z      		&		 \citet{2011ApJS..193...24S}  	\\	
HD 46056		&		46056		&		8.16			&		2		&		0		&		0		&		2			&		O8		&		V		&		n		&		\citet{2011ApJS..193...24S}	\\	
 HD 46106    		&		46106		&		7.948	 		&		1		&		0		&		0		&		1			&		O9.7		&		II-III		&				&		\citet{2011ApJS..193...24S}	\\	
 HD 46149     		&		46149		&		7.59	 		&		6		&		0		&		0		&		6			&		 O8.5  		&		 V       		&		             		&		 \citet{2011ApJS..193...24S}  	\\	
 HD 46150     		&		46150		&		6.75	 		&		8		&		0		&		0		&		8			&		 O5    		&		 V       		&		 ((f))z      		&		 \citet{2011ApJS..193...24S}  	\\	
 HD 46202     		&		46202		&		8.2	 		&		3		&		0		&		0		&		3			&		 O9.5  		&		 V       		&		             		&		 \citet{2011ApJS..193...24S}  	\\	
 HD 46223     		&		46223		&		7.32	 		&		5		&		0		&		0		&		5			&		 O4    		&		 V       		&		 ((f))       		&		 \citet{2011ApJS..193...24S}  	\\	
 HD 46485     		&		46485		&		8.2	 		&		5		&		0		&		0		&		5			&		 O7    		&		 V       		&		 n           		&		 \citet{2011ApJS..193...24S}  	\\	
 HD 46966     		&		46966		&		6.87	 		&		1		&		0		&		0		&		1			&		 O8.5  		&		 IV      		&		             		&		 \citet{2011ApJS..193...24S}  	\\	
 V640 Mon     		&		47129		&		6.06	 		&		38		&		4		&		0		&		42			&		 O8    		&		         		&		 fp var      		&		 \citet{2011ApJS..193...24S}  	\\	
 V689 Mon     		&		47432		&		6.24	 		&		1		&		0		&		0		&		1			&		 O9.7  		&		 Ib      		&		             		&		 \citet{2011ApJS..193...24S}  	\\	
 15 Mon       		&		47839		&		4.64	 		&		3		&		13		&		0		&		16			&		 O7    		&		 V       		&		 ((f)) var   		&		 \citet{2011ApJS..193...24S}  	\\	
 HD 48099     		&		48099		&		6.37	 		&		2		&		1		&		0		&		3			&		 O6.5  		&		 V       		&		 (n)((f))    		&		 \citet{2011ApJS..193...24S}  	\\	
 HD 54662     		&		54662		&		6.23	 		&		2		&		0		&		0		&		2			&		 O7    		&		 V       		&		 ((f))z var? 		&		 \citet{2011ApJS..193...24S}  	\\	
HD 55879    		&		55879		&		6.0	 		&		1		&		0		&		0		&		1			&		O9.7		&		III		&				&		\citet{2011ApJS..193...24S}	\\	
HD 57682     		&		57682		&		6.4	 		&		37		&		0		&		0		&		37			&		 O9.5  		&		 IV      		&		             		&		 \citet{2011ApJS..193...24S}  	\\	
 HD 66788     		&		66788		&		9.43	 		&		2		&		0		&		0		&		2			&		 O8    		&		 V       		&		             		&		\citet{2003ApJS..144...21R}	\\	
 $\zeta$ Pup         		&		66811		&		2.25	 		&		30		&		0		&		5		&		35			&		 O4    		&		 I       		&		 f           		&		\citet{2013msao.confE.101S}	\\	
WR 11		&		68273		&		1.83	 		&		12		&		0		&		0		&			12		&		WC8+O7.5III-V 		&				&				&		\citet{crowther}	\\	
 HD 69106    		&		69106		&		7.13	 		&		1		&		0		&		0		&		1			&		O9.7		&		In		&				&		\citet{2013msao.confE.101S}	\\	
 HD 93028    		&		93028		&		8.36	 		&		0		&		0		&		2		&		2			&		 O9    		&		 IV       		&		             		&		\citet{2011ApJS..193...24S}	\\	
 HD 93250    		&		93250		&		7.5	 		&		0		&		0		&		2		&		2			&		 O4    		&		 III     		&		 fc:         		&		\citet{2011ApJS..193...24S}	\\	
HD 148937    		&		148937		&		6.77	 		&		32		&		0		&		0		&		32			&		O6		&		         		&		 f?p         		&		 \citet{2013msao.confE.101S} 	\\	
$\mu$ Nor		&		149038		&		4.914	 		&		3		&		0		&		0		&		3			&		O9.7		&		Iab		&				&		\citet{2011ApJS..193...24S}	\\	
 $\zeta$ Oph      		&		149757		&		2.58	 		&		20		&		46		&		0		&		66			&		 O9.5  		&		 IV      		&		 nn          		&		 \citet{2011ApJS..193...24S}  	\\	
 V973 Sco                		&		151804		&		5.249	 		&		0		&		0		&		2		&		2			&		O8		&		Ia			&			&	\citet{2011ApJS..193...24S}	\\	
 HD 152233    		&		152233		&		6.59	 		&		1		&		0		&		0		&		1			&		 O6    		&		 Ib      		&		 (f)         		&		\citet{2013msao.confE.101S}	\\	
 HD 152247    		&		152247		&		7.2	 		&		1		&		0		&		0		&		1			&		 O9.5  		&		 III     		&		             		&		\citet{2013msao.confE.101S}	\\	
 HD 152249    		&		152249		&		6.47	 		&		1		&		0		&		0		&		1			&		 OC9   		&		 Iab     		&		             		&		 \citet{2011ApJS..193...24S}  	\\	
 HD 152408            &             152408           &             5.77                 &             3              &		0		&		0		&		1			&		 O9   		&		 II-III     		&		             		&		\citet{2013msao.confE.101S}  	\\	
 HD 153426    		&		153426		&		7.49	 		&		2		&		0		&		0		&		2			&		 O9    		&		 II-III  		&		             		&		\citet{2013msao.confE.101S}	\\	
 V884 Sco     		&		153919		&		6.53	 		&		1		&		0		&		0		&		1			&		 O6    		&		 Ia      		&		 f           		&		\citet{2013msao.confE.101S}	\\	
 V1074 Sco    		&		154368		&		6.18	 		&		1		&		0		&		0		&		1			&		 O9.5  		&		 Iab     		&		             		&		 \citet{2011ApJS..193...24S}  	\\	
HD 154643		&		154643		&		7.15	 		&		1		&		0		&		0		&		1			&		 O9.5  		&		 III     		&		             		&		\citet{2011ApJS..193...24S}	\\	
  \hline																																																
\end{tabular}																																																
\end{center}																																																
\end{table*}																																																
																																																
\begin{table*}																																																
\contcaption{MiMeS O-type and WR SC targets.}																																																
\begin{center}\begin{tabular}{llllllllllllllrrc}																																																
\hline																						
& & & \multicolumn{4}{c}{\# of observations}\\									
Name		&		HD		&		V			&		E		&		N		&		H		&		Tot		&	Spectral type		&		LC		&		pec		&		Reference	\\	
\hline                 																																																
HD 155806		&		155806		&		5.612	 		&		25		&		0		&		0		&		25			&		O7.5		&		V		&		 		&		\citet{2013msao.confE.101S}	\\	
HD 155889		&		155889		&		6.565	 		&		0		&		0		&		1		&		1			&		O9.5		&		IV		&				&		\citet{2011ApJS..193...24S}	\\	
 HD 156154    		&		156154		&		8.04	 		&		1		&		0		&		0		&		1			&		 O7.5  		&		 Ib      		&		             		&		 \citet{2011ApJS..193...24S}  	\\	
 63 Oph       		&		162978		&		6.2	 		&		5		&		0		&		0		&		5			&		 O8    		&		 II      		&		 ((f))       		&		 \citet{2011ApJS..193...24S}  	\\	
 HD 164492    		&		164492		&		7.53	 		&		1		&		0		&		0		&		1			&		 O7.5  		&		 V       		&		 z           		&		\citet{2013msao.confE.101S}	\\	
 9 Sgr        		&		164794		&		5.93	 		&		14		&		3		&		0		&		17			&		 O4    		&		 V       		&		 ((fc))      		&		\citet{2013msao.confE.101S}	\\	
 HD 165052    		&		165052		&		6.84	 		&		1		&		0		&		0		&		1			&		 O5.5:+O8:  		&		 Vz+V 		&		             		&		\citet{2011ApJS..193...24S}	\\	
WR 111		&		165763		&		7.82	 		&		16		&		0		&		0		&			16		&		WC5		&				&				&		\citet{crowther}	\\	
16 Sgr		&		167263		&		5.97	 		&		1		&		0		&		0		&		1			&		 O9.5 		&		 II-III               		&		 n 		&		\citet{2013msao.confE.101S}	\\	
15 Sgr 		&		167264		&		5.347	 		&		5		&		7		&		0		&		12			&		O9.7		&		Iab		&				&		\citet{2013msao.confE.101S}	\\	
 HD 167771    		&		167771		&		6.54	 		&		2		&		0		&		0		&		2			&		 O7    		&		 III     		&		 (f)         		&		 \citet{2011ApJS..193...24S}  	\\	
 HD 186980    		&		186980		&		7.48	 		&		1		&		0		&		0		&		1			&		 O7.5  		&		 III     		&		 ((f))       		&		\citet{2013msao.confE.101S}	\\	
 9 Sge        		&		188001		&		6.24	 		&		1		&		0		&		0		&		1			&		 O7.5  		&		 Iab     		&		 f           		&		 \citet{2011ApJS..193...24S}  	\\	
 HD 188209    		&		188209		&		5.63	 		&		1		&		27		&		0		&		28			&		 O9.5  		&		 Iab     		&		             		&		 \citet{2011ApJS..193...24S}  	\\	
HD 189957		&		189957		&		7.82	 		&		1		&		0		&		0		&		1			&		 O9.7		&		 III     		&		             		&		 \citet{2013msao.confE.101S}  	\\	
WR 133		&		190918		&		6.75	 		&		3		&		0		&		0		&			3		&		WN5o+O9I  		&				&				&		\citet{crowther}	\\	
 HD 191201    		&		191201		&		7.34	 		&		1		&		0		&		0		&		1			&		 O9.5+B0		&		 III+IV 		&		 ((n))       		&		\citet{2011ApJS..193...24S}	\\	
WR 134		&		191765		&		8.08	 		&	 38		&		0		&		0		&			38		&		WN6b		&				&				&		\citet{crowther}	\\	
WR 135		&		192103		&		8.11	 		&		13		&		0		&		0		&		13			&		WC8		&				&				&		\citet{crowther}	\\	
WR 136		&		192163		&		7.5	 		&		9		&		0		&		0		&			9		&		WN6b		&				&		(h)		&		\citet{crowther}	\\	
 V2011 Cyg    		&		192281		&		7.55	 		&		0		&		2		&		0		&		2			&		 O4.5  		&		 V       		&		 n(f)        		&		 \citet{2011ApJS..193...24S}  	\\	
 HD 192639    		&		192639		&		7.11	 		&		0		&		1		&		0		&		1			&		 O7.5  		&		 Iab     		&		 f           		&		 \citet{2011ApJS..193...24S}  	\\	
WR 137		&		192641		&		7.91	 		&		27		&		0		&		0		&		27			&		WC7pd+O9		&				&				&		\citet{crowther}	\\	
WR 138		&		193077		&		8.01	 		&		6		&		0		&		0		&			6		&		WN6o		&				&				&		\citet{crowther}	\\	
 HD 193322    		&		193322		&		5.82	 		&		2		&		0		&		0		&		2			&		 O9    		&		 IV      		&		 (n)         		&		 \citet{2011ApJS..193...24S}  	\\	
 HD 193443    		&		193443		&		7.24	 		&		1		&		0		&		0		&		1			&		 O9    		&		 III     		&		             		&		 \citet{2011ApJS..193...24S}  	\\	
WR 139		&		193576		&		8.00	 		&		9		&		0		&		0		&		9			&		WN5o+O6III-V  		&				&				&		\citet{crowther}	\\	
WR 140		&		193793		&		6.85	 		&		19		&		0		&		0		&	19		&		WC7pd+O4-5 		&				&				&		\citet{crowther}	\\	
 HD 199579    		&		199579		&		5.97	 		&		4		&		0		&		0		&		4			&		 O6.5  		&		 V       		&		 ((f))z      		&		 \citet{2011ApJS..193...24S}  	\\	
 HD 201345    		&		201345		&		7.75	 		&		1		&		0		&		0		&		1			&		 ON9.5 		&		 IV      		&		             		&		 \citet{2011ApJS..193...24S}  	\\	
 68 Cyg       		&		203064		&		5.04	 		&		5		&		3		&		0		&		8			&		 O7.5  		&		 III     		&		 n((f))      		&		 \citet{2011ApJS..193...24S}  	\\	
 HD 204827    		&		204827		&		8.00	 		&		1		&		0		&		0		&		1			&		 O9.7  		&		 III     		&		             		&		 \citet{2011ApJS..193...24S}  	\\	
 HD 206183    		&		206183		&		7.41	 		&		1		&		0		&		0		&		1			&		 O9.5  		&		 IV-V    		&		             		&		 \citet{2011ApJS..193...24S}  	\\	
 HD 206267    		&		206267		&		5.62	 		&		6		&		0		&		0		&		6			&		 O6.5  		&		 V       		&		 ((f))       		&		 \citet{2011ApJS..193...24S}  	\\	
 HD 207198    		&		207198		&		5.96	 		&		1		&		15		&		0		&		16			&		 O9    		&		 II      		&		             		&		 \citet{2011ApJS..193...24S}  	\\	
 HD 207538    		&		207538		&		7.3	 		&		1		&		0		&		0		&		1			&		 O9.7  		&		 IV      		&		             		&		 \citet{2011ApJS..193...24S}  	\\	
 14 Cep       		&		209481		&		5.55	 		&		3		&		11		&		0		&		14			&		 O9    		&		 IV      		&		 (n) var     		&		 \citet{2011ApJS..193...24S}  	\\	
 19 Cep       		&		209975		&		5.11	 		&		7		&		26		&		0		&		33			&		 O9    		&		 Ib      		&		             		&		 \citet{2011ApJS..193...24S}  	\\	
 HD 210809    		&		210809		&		7.56	 		&		1		&		0		&		0		&		1			&		 O9    		&		 Iab     		&		             		&		 \citet{2011ApJS..193...24S}  	\\	
 $\lambda$ Cep      		&		210839		&		5.08	 		&		0		&		26		&		0		&		26			&		 O6.5  		&		 I       		&		 (n)fp       		&		 \citet{2011ApJS..193...24S}  	\\	
 10 Lac       		&		214680		&		4.88	 		&		1		&		35		&		0		&		36			&		 O9    		&		 V       		&		             		&		 \citet{2011ApJS..193...24S}  	\\	
 HD 218195    		&		218195		&		8.44	 		&		1		&		0		&		0		&		1			&		 O8.5  		&		 III     		&		             		&		 \citet{2011ApJS..193...24S}  	\\	
 HD 218915    		&		218915		&		7.2	 		&		1		&		0		&		0		&		1			&		 O9.5  		&		 Iab     		&		             		&		 \citet{2011ApJS..193...24S}  	\\	
 HD 227757    		&		227757		&		9.27	 		&		2		&		0		&		0		&		2			&		 O9.5  		&		 V       		&		             		&		Maiz Apellaniz (priv. comm.)	\\	
 HD 258691    		&		258691		&		9.7	 		&		3		&		0		&		0		&		3			&		 O9.5  		&		 IV      		&		             		&		Maiz Apellaniz (priv. comm.)	\\	
 HD 328856    		&		328856		&		8.5	 		&		1		&		0		&		0		&		1			&		 O9.7  		&		 Ib      		&		             		&		\citet{2013msao.confE.101S}	\\	
 BD+60 499    		&		            		&		10.27	 		&		1		&		0		&		0		&		1			&		 O9.5  		&		 V       		&		             		&		 \citet{2011ApJS..193...24S}  	\\	
CD-28 5104   		&		            		&		10.09	 		&		44		&		0		&		2		&		46			&		 O6.5  		&		         		&		 f?p         		&		\citet{2013msao.confE.101S}	\\	
 BD-13 4930   		&		            		&		9.37	 		&		1		&		0		&		0		&		1			&		 O9.7  		&		 V       		&		             		&		Maiz Apellaniz (priv. comm.)	\\	
NGC 1624 2   		&		            		&		11.77	 		&		17		&		1		&		0		&		18			&		 O7    		&		         		&		 f?p         		&		 \citet{2011ApJS..193...24S}  	\\	
 \hline																																																
 \end{tabular}																																																
   \end{center}																																																
\end{table*}


\begin{table*}																						
\caption{\label{SCB}MiMeS B-type SC targets. Columns report common identifier and HD \#, $V$ band magnitude, the number of ESPaDOnS (E), Narval (N) and HARPSpol (H) spectra acquired, and the total number of spectra, the spectral type, luminosity class (LC) and any spectral peculiarity (pec), and the reference to the spectral type, luminosity class and peculiarity. 'BSC' indicates the Bright Star Catalogue \citep{1991bsc..book.....H}. }																																																
\begin{center}
\begin{tabular}{llllllllllllllrrc}																																																
\hline																																																& & & \multicolumn{4}{c}{\# of observations}\\									
	Name		&		HD		&		V		&		E		&		N		&		H		&		Tot			&	Spectral type		&		LC		&		pec		&		Reference	\\	
\hline																																																
 $\alpha$ And                		&		358		&		2.058	 		&		16		&		0		&		0		&		16		&		B8	&	IV 	&	 p MnHg		&		BSC	\\							
 $\gamma$ Peg                  		&		886		&		2.83	 		&		6		&		23		&		0		&		29		&		B2	&	IV		&		&		BSC	\\							
 HD 955                   		&		955		&		7.369	 		&		0		&		0		&		2		&		2		&		B4	&	V	&			&		\citet{2003AJ....125.2531R}	\\							
 V746 Cas                 		&		1976		&		5.575	 		&		0		&		16		&		0		&		16		&		B5	&	IV	&			&		BSC	\\							
 $\kappa$ Cas                  		&		2905		&		4.189	 		&		3		&		0		&		0		&		3		&		B1	&	Ia	&	e		&		BSC	\\							
 53 Psc                   		&		3379		&		5.864	 		&		1		&		10		&		0		&		11		&		B2.5	&	IV	&			&		BSC	\\
\hline																																																
\end{tabular}																																																
\end{center}		

\end{table*}																																																
																																																
\begin{table*}																						
\contcaption{MiMeS B-type SC targets.}																																																
\begin{center}\begin{tabular}{llllllllllllllrrc}																																																
\hline																																							& & & \multicolumn{4}{c}{\# of observations}\\																		
Name		&		HD		&		V		&		E		&		N		&		H		&		Tot		&	Spectral type		&		LC		&		pec		&		Reference	\\		
\hline                 																										
HD 36629    &             				36629		&		7.64	 		&		1		&		0		&			1		&		B2.5	&	IV		&		&	&	\citet{1977PASP...89..797A} 	\\							
 37 Ori                   		&		36822		&		4.398	 		&		1		&		0		&		0		&		1		&		B0	&	III		&		&		BSC	\\							
 HD 36959                 		&		36959		&		5.67	 		&		1		&		0		&		0		&		1		&		B1	&	V	&	v		&		BSC	\\							
 HD 36960                 		&		36960		&		4.78	 		&		1		&		0		&		0		&		1		&		B0.5	&	V		&		&		BSC	\\							
 V1046 Ori                		&		37017		&		6.553	 		&		10		&		0		&		0		&		10		&		B1.5	&	V		&		&		BSC	\\							
 41 Ori A                 		&		37020		&		6.73	 		&		3		&		0		&		0		&		3		&	   B0.2  		&		    V   	&	        		&		            	Maiz Apellaniz (priv. comm.)	  \\
\hline																																																
\end{tabular}																																																
\end{center}																																																
\end{table*}																																																
																																																
\begin{table*}																						
\contcaption{MiMeS B-type SC targets.}																																																
\begin{center}\begin{tabular}{llllllllllllllrrc}																																																
\hline																																																
& & & \multicolumn{4}{c}{\# of observations}\\									
Name		&		HD		&		V		&		E		&		N		&		H		&		Tot		&	Spectral type		&		LC		&		pec		&		Reference	\\		
\hline                 																																																
HD 58647                 		&		58647		&		6.85	 		&		0		&		0		&		1		&		1		&		B9	&	IV	&	e		&		\citet{1988mcts.book.....H}	\\					
 $\beta$ CMi                  		&		58715		&		2.886	 		&		20		&		14		&		0		&		34		&		B8	&	V	&	e		&		BSC	\\							
 FY CMa                   		&		58978		&		5.56	 		&		3		&		0		&		0		&		3		&		B0	&	IV	&	:pe		&		BSC	\\							
 V343 Pup                     	&	60431	&	     8.57       		&	1	&		0		&		0		&		1		&	    B4 	&	  	&	 p SiMg 	&	 {\citet{2009AA...498..961R}}  \\ 	
 PT Pup                   		&		61068		&		5.693	 		&		1		&		0		&		0		&		1		&		B2	&	III	&			&		BSC	\\							
 HD 61555                 		&		61555		&		4.635	 		&		3		&		0		&		0		&		3		&		B6	&	V	&			&		BSC	\\
\hline																																																
\end{tabular}																																																
\end{center}																																																
\end{table*}																																																
																																																
\begin{table*}																							
\contcaption{MiMeS B-type SC targets.}																																																
\begin{center}\begin{tabular}{llllllllllllllrrc}																																																
\hline																																																
& & & \multicolumn{4}{c}{\# of observations}\\									
Name		&		HD		&		V		&		E		&		N		&		H		&		Tot		&	Spectral type		&		LC		&		pec		&		Reference	\\		
\hline                 																																																
 V507 Car                 		&		90966		&		6.429	 		&		0		&		0		&		2		&		2		&		B2/B3	&	III	&	:ne		&		\citet{1978MSS...C02....0H}	\\							
 HD 91120                 		&		91120		&		5.592	 		&		0		&		4		&		0		&		4		&		B9	&	V	&	ne		&		BSC	\\							
 $\rho$ Leo                  		&		91316		&		3.87	 		&		2		&		5		&		0		&		7		&		B1	&	Ib	&			&		BSC	\\							
 HD 91969                 		&		91969		&		6.51	 		&		0		&		0		&		1		&		1		&		B0	&	Iab	&		e	&		\citet{2005AA...437..467E}	\\							
 V518 Car                 		&		92938		&		4.82	 		&		0		&		0		&		1		&		1		&		B4	&	V	&			&		BSC	\\							
 $\theta$ Car                  		&		93030		&		2.78	 		&		0		&		0		&		3		&		3		&		B0	&	V	&	p		&		BSC	\\
\hline																																																
\end{tabular}																																																
\end{center}																																																
\end{table*}																																																
																																																
\begin{table*}																							
\contcaption{MiMeS B-type SC targets.}																																																
\begin{center}\begin{tabular}{llllllllllllllrrc}																																																
\hline																																																
& & & \multicolumn{4}{c}{\# of observations}\\									
Name		&		HD		&		V		&		E		&		N		&		H		&		Tot		&	Spectral type		&		LC		&		pec		&		Reference	\\		
\hline                 																																																
 $\alpha$ Lup                  		&		129056		&		2.276	 		&		0		&		0		&		1		&		1		&		B1.5	&	III		&		&		BSC	\\							
 V836 Cen                 		&		129929		&		8.05	 		&		1		&		0		&		0		&		1		&		B3	&	V		&		&		\citet{1970MNRAS.150...23H}	\\							
  $o$ Lup                  		&		130807		&		4.313	 		&		0		&		0		&		12		&		12		&		B5	&	IV	&			&		BSC	\\							
 $\beta$ Lup                  		&		132058		&		2.665	 		&		0		&		0		&		6		&		6		&		B2	&	III		&		&		BSC	\\							
 $\kappa$ Cen                  		&		132200		&		3.13	 		&		0		&		0		&		2		&		2		&		B2	&	IV		&		&		BSC	\\							
 HD 132955                		&		132955		&		5.427	 		&		0		&		0		&		1		&		1		&		B3	&	V	&			&		BSC	\\
\hline																																																
\end{tabular}																																																
\end{center}																																																
\end{table*}																																																
																																																
\begin{table*}																						
\contcaption{MiMeS B-type SC targets.}																																																
\begin{center}\begin{tabular}{llllllllllllllrrc}																																																
\hline																																																
& & & \multicolumn{4}{c}{\# of observations}\\									
Name		&		HD		&		V		&		E		&		N		&		H		&		Tot		&	Spectral type		&		LC		&		pec		&		Reference	\\		
\hline                 																																																
 $\kappa$ Sco                  		&		160578		&		2.375	 		&		0		&		0		&		2		&		2		&		B1.5	&	III	&			&		BSC	\\							
 $\iota$ Her                  		&		160762		&		3.8	 		&		0		&		3		&		0		&		3		&		B3	&	IV		&		&		BSC	\\							
 V1092 Sco                		&		163254		&		6.74	 		&		3		&		0		&		0		&		3		&		B2	&	V	&	SB2		&		\citet{1977ApJS...35..111G}	\\							
 66 Oph                   		&		164284		&		4.784	 		&		24		&		32		&		0		&		56		&		B2	&	V	&	e		&		BSC	\\							
 HD 164402                		&		164402		&		5.749	 		&		2		&		0		&		0		&		2		&		B0	&	Ib		&		&		BSC	\\							
 HD 164865                		&		164865		&		7.74	 		&		0		&		0		&		2		&		2		&		B9	&	Ia		&		&		\citet{1977ApJS...35..111G}	\\
\hline																											
\end{tabular}																																																
\end{center}																																																
\end{table*}																																																
																																																
\begin{table*}																						
\contcaption{MiMeS B-type SC targets.}																																																
\begin{center}\begin{tabular}{llllllllllllllrrc}																																																
\hline																																																
& & & \multicolumn{4}{c}{\# of observations}\\									
Name		&		HD		&		V		&		E		&		N		&		H		&		Tot		&	Spectral type		&		LC		&		pec		&		Reference	\\		
\hline                 																																																
 HD 206540                		&		206540		&		6.052	 		&		1		&		2		&		0		&		3		&		B5	&	IV	&			&		BSC	\\							
 81 Cyg                   		&		207330		&		4.18	 		&		0		&		2		&		0		&		2		&		B3	&	III		&		&		BSC	\\							
 HD 208213                		&		208213		&		8.54	 		&		0		&		0		&		2		&		2		&		B3	&	IV	&			&	\citet{1982mcts.book.....H}		\\							
 $o$ Aqr                  		&		209409		&		4.7	 		&		0		&		2		&		0		&		2		&		B7	&	IV	&	e		&		BSC	\\							
 35 Aqr                   		&		210191		&		5.791	 		&		0		&		0		&		1		&		1		&		B2.5	&	IV	&			&		BSC	\\							
 31 Peg                   		&		212076		&		4.81	 		&		24		&		0		&		0		&		24		&		B2	&	IV-V	&	e		&		BSC	\\
\hline																																																
\end{tabular}																																																
\end{center}																																																
\end{table*}																																																
																																																
\begin{table*}																						
\contcaption{MiMeS B-type SC targets.}																																																
\begin{center}\begin{tabular}{llllllllllllllrrc}																																																
\hline																																																
& & & \multicolumn{4}{c}{\# of observations}\\									
Name		&		HD		&		V		&		E		&		N		&		H		&		Tot		&	Spectral type		&		LC		&		pec		&		Reference	\\		
\hline                 																																																
 BD+04 1299s              		&		             		&		9.735	 		&		0		&		1		&		0		&		1		&		B1	&	III		&		&		\citet{1995ApJ...454..151M}\\							
 BD+05 1281B              		&		             		&		10.73	 		&		1		&		0		&		0		&		1		&		B1.5	&	V		&		&		\citet{1995ApJ...454..151M}	\\							
 BD+05 1283E              		&		             		&		11.813	 		&		1		&		0		&		0		&		1		&		B5	&	Iab		&		&	\citet{1965ApJ...142..974M}		\\							
 BD+09 1331B              		&		             		&		10.79	 		&		1		&		0		&		0		&		1		&		B1	&	V		&		&		\citet{2001AA...373..625P}	\\							
 BD+10 1222               		&		             		&		9.88	 		&		1		&		0		&		0		&		1		&		B8		&	V	&		&		\citet{1987PASP...99.1050P} 	\\							
 BD+41 3731               		&		             		&		9.84	 		&		1		&		1		&		0		&		2		&		B2		&		&		&		\citet{1985AA...151..340F}	\\
\hline																																												\end{tabular}																																																
   \end{center}																																																
\end{table*}

\clearpage
\begin{table*}
\begin{flushleft}
{\Huge Affiliations}\\
\ \\
$^{1}$Department of Physics, Royal Military College of Canada, PO Box 17000, Stn Forces, Kingston, Ontario K7K 7B4, Canada \\
$^2$LESIA, Observatoire de Paris, PSL Research University, CNRS, Sorbonne Universit\'es, UPMC, Univ. Paris 6, Univ. Paris Diderot, Sorbonne Paris Cit\'e, 5 place Jules Janssen, 92195, Meudon, France\\
$^3$UJF-Grenoble 1/CNRS-INSU, Institut de Plan\'etologie et d'Astrophysique de Grenoble (IPAG) UMR 5274, 38041, Grenoble, France\\
$^{4}$CNRS-IPAG, F-38000, Grenoble, France\\
$^5$European Organisation for Astronomical Research in the Southern Hemisphere, Karl-Schwarzschild-Str. 2, D-85748 Garching, Germany\\
$^6$Dept. of Physics \& Space Sciences, Florida Institute of Technology, Melbourne, FL, USA\\
$^{7}$Herzberg Astronomy and Astrophysics Program, National Research Council of Canada, 5071 West Saanich Road, Victoria, BC V9E 2E7, Canada\\
$^{8}$Swarthmore College, Department of Physics and Astronomy, Swarthmore, PA 19081, USA\\
$^{9}$Anton Pannekoek Institute for Astronomy, University of Amsterdam, Science Park 904, 1098 XH Amsterdam, The Netherlands\\
$^{10}$Department of Physics and Astronomy, Uppsala University, Box 516, 75120, Uppsala, Sweden\\
$^{11}$Physics \& Astronomy Department, The University of Western Ontario, London, Ontario, N6A 3K7, Canada\\
$^{12}$Armagh Observatory, College Hill, Armagh BT61 9DG, UK\\
$^{13}$Canada-France-Hawaii Telescope Corporation, 65-1238 Mamalahoa Hwy Kamuela HI 96743, USA\\
$^{14}$LUPM, CNRS \& Universit\'e de Montpellier, Place EugËne Bataillon, F-34095, Montpellier Cedex 05, France\\
$^{15}$Laboratoire AIM Paris-Saclay, CEA/DSM, CNRS, Universit\'e Paris Diderot, IRFU/SAp Centre de Saclay, F-91191 Gif-sur-Yvette, France\\
$^{16}$University of Delaware, Bartol Research Institute, Newark, DE 19716, USA\\
$^{17}$European Organisation for Astronomical Research in the Southern Hemisphere, Casilla 19001, Santiago, Chile\\
$^{18}$Department of Physics, Engineering Physics and Astronomy, Queen's University, 99 University Avenue, Kingston, ON K7L 3N6, Canada\\
$^{19}$Institut f\"ur Astronomie und Astrophysik der Universit\"at M\"unchen, Scheinerstr. 1, D-81679 M\"unchen, Germany\\
$^{20}$Department of Astronomy, University of Wisconsin Madison, 2535 Sterling Hall, 475 North Charter St, Madison, WI 53706, USA\\
$^{21}$Department of Physics, Penn State Worthington Scranton, 120 Ridge View Drive, Dunmore, PA 18512, USA\\
$^{22}$Aix Marseille Universit\'e, CNRS, LAM (Laboratoire d'Astrophysique de Marseille) UMR 7326, 13388, Marseille, France\\
$^{23}$Argelander Institut f\"ur Astronomie, Auf dem H\"ugel 71, 53121 Bonn, Germany\\
$^{24}$Institut d?Astrophysique et de G\'eophysique, Universit\'e de Li\`ege, Quartier Agora, 19c, All\'ee du 6 Ao\^ut, B5c, B-4000 Sart Tilman, Belgium\\
$^{25}$Instituto de Astronomia, Geof\'{i}sica e Ci\^{e}ncias Atmosf\'{e}ricas, Universidade de S\~{a}o Paulo, Rua do Mat\~{a}o 1226, Cidade Universit\'{a}ria, S\~{a}o Paulo, SP 05508-900, Brazil\\
$^{26}$Space Telescope Science Institute, 3700 San Martin Drive, Baltimore, MD 21218, USA\\
$^{27}$Universidade Federal do Rio de Janeiro, Observat\'orio do Valongo, Ladeira Pedro AntÙnio, 43, CEP 20.080-090, Rio de Janeiro, Brazil\\
$^{28}$D\'epartement de Physique and Centre de Recherche en Astrophysique du QuÈbec, Universit\'e de Montr\'eal, C.P. 6128, Succ. C-V, Montreal, QC, H3C 3J7, Canada\\
$^{29}$Universit\'e de Toulouse; UPS-OMP; IRAP; Toulouse, France\\
$^{30}$CNRS-IRAP; 14, avenue Edouard Belin, 31400 Toulouse, France\label{inst:irap2}\\
$^{31}$Max Planck Insitut f\"ur Extraterrestrische Physik, Giessenbachstrasse 1, 85748 Garching, Germany\\
$^{32}$Departamento de F\'{\i}sica, Universidad de La Serena, Av. Juan Cisternas 1200 Norte, La Serena, Chile\\
$^{33}$Observatoire de Paris, PSL Research University, 61 avenue de l'Observatoire, 75014 Paris, France\\
$^{34}$Department of Mathematics, Mathematical Sciences Institute, The Australian National University, ACT 0200 Australia\\
$^{35}$Kiepenheuer-Institut f\"{u}r Sonnenphysik, Sch\"{o}neckstr. 6, D-79104 Freiburg, Germany \\
$^{36}$Institute for Astronomy, University of Hawaii, 2680 Woodlawn Drive, Honolulu, HI, 96822, USA  \\
$^{37}$Applied Research Labs, University of Hawaii, 2800 Woodlawn Drive, Honolulu, HI, 96822, USA \\
$^{38}$Department of Physics \& Astronomy, University College London, Gower St, London WC1E 6BT, UK\\
$^{39}$Department of Physics and Astronomy, East Tennessee State University, Johnson City, TN 37614, USA\\
$^{40}$Department of Astrophysics, University of Vienna, Tuerkenschanzstrasse 17, 1180 Vienna, Austria\\
$^{41}$Special Astrophysical Observatory, Russian Academy of Sciences, Nizhny Arkhyz, Russia 369167\\
$^{42}$Centro de Astrobiolog{\'i}a, CSIC-INTA,\ Ctra. Torrej{\'o}n a Ajalvir km.4, 28850 Madrid, Spain \\
$^{43}$Instituut voor Sterrenkunde, KU Leuven, Celestijnenlaan 200D, 3001 Leuven, Belgium
\end{flushleft}
\end{table*}

\end{document}